\documentclass[twoside,twocolumn,9pt]{article}
\usepackage{extsizes}
\usepackage[super,sort&compress,comma]{natbib} 
\usepackage[version=3]{mhchem}
\usepackage[left=1.5cm, right=1.5cm, top=1.785cm, bottom=2.0cm]{geometry}
\usepackage{balance}
\usepackage{mathptmx}
\usepackage{sectsty}
\usepackage{graphicx} 
\usepackage{lastpage}
\usepackage[format=plain,justification=justified,singlelinecheck=false,font={stretch=1.125,small,sf},labelfont=bf,labelsep=space]{caption}
\usepackage{float}
\usepackage{fancyhdr}
\usepackage{fnpos}
\usepackage[english]{babel}
\addto{\captionsenglish}{%
  
}
\usepackage{array}
\usepackage{droidsans}
\usepackage{charter}
\usepackage[T1]{fontenc}
\usepackage[usenames,dvipsnames]{xcolor}
\usepackage{setspace}
\usepackage[compact]{titlesec}
\usepackage{hyperref}

\usepackage{epstopdf}

\definecolor{cream}{RGB}{222,217,201}

\begin{document}

\pagestyle{fancy}
\thispagestyle{plain}
\fancypagestyle{plain}{
\renewcommand{\headrulewidth}{0pt}
}

\makeFNbottom
\makeatletter
\renewcommand\LARGE{\@setfontsize\LARGE{15pt}{17}}
\renewcommand\Large{\@setfontsize\Large{12pt}{14}}
\renewcommand\large{\@setfontsize\large{10pt}{12}}
\renewcommand\footnotesize{\@setfontsize\footnotesize{7pt}{10}}
\makeatother

\renewcommand{\thefootnote}{\fnsymbol{footnote}}
\renewcommand\footnoterule{\vspace*{1pt}%
\color{cream}\hrule width 3.5in height 0.4pt \color{black}\vspace*{5pt}} 
\setcounter{secnumdepth}{5}

\makeatletter 
\renewcommand\@biblabel[1]{#1}            
\renewcommand\@makefntext[1]%
{\noindent\makebox[0pt][r]{\@thefnmark\,}#1}
\makeatother 
\renewcommand{\figurename}{\small{Fig.}~}
\sectionfont{\sffamily\Large}
\subsectionfont{\normalsize}
\subsubsectionfont{\bf}
\setstretch{1.125} 
\setlength{\skip\footins}{0.8cm}
\setlength{\footnotesep}{0.25cm}
\setlength{\jot}{10pt}
\titlespacing*{\section}{0pt}{4pt}{4pt}
\titlespacing*{\subsection}{0pt}{15pt}{1pt}

\fancyfoot{}
\fancyfoot[LO,RE]{\vspace{-7.1pt}\includegraphics[height=9pt]{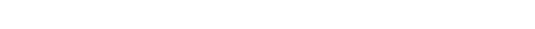}}
\fancyfoot[CO]{\vspace{-7.1pt}\hspace{13.2cm}\includegraphics{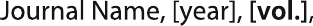}}
\fancyfoot[CE]{\vspace{-7.2pt}\hspace{-14.2cm}\includegraphics{head_foot/RF}}
\fancyfoot[RO]{\footnotesize{\sffamily{1--\pageref{LastPage} ~\textbar  \hspace{2pt}\thepage}}}
\fancyfoot[LE]{\footnotesize{\sffamily{\thepage~\textbar\hspace{3.45cm} 1--\pageref{LastPage}}}}
\fancyhead{}
\renewcommand{\headrulewidth}{0pt} 
\renewcommand{\footrulewidth}{0pt}
\setlength{\arrayrulewidth}{1pt}
\setlength{\columnsep}{6.5mm}
\setlength\bibsep{1pt}

\makeatletter 
\newlength{\figrulesep} 
\setlength{\figrulesep}{0.5\textfloatsep} 

\newcommand{\topfigrule}{\vspace*{-1pt}%
\noindent{\color{cream}\rule[-\figrulesep]{\columnwidth}{1.5pt}} }

\newcommand{\botfigrule}{\vspace*{-2pt}%
\noindent{\color{cream}\rule[\figrulesep]{\columnwidth}{1.5pt}} }

\newcommand{\dblfigrule}{\vspace*{-1pt}%
\noindent{\color{cream}\rule[-\figrulesep]{\textwidth}{1.5pt}} }

\makeatother

\twocolumn[
\begin{@twocolumnfalse}
	{\includegraphics[height=30pt]{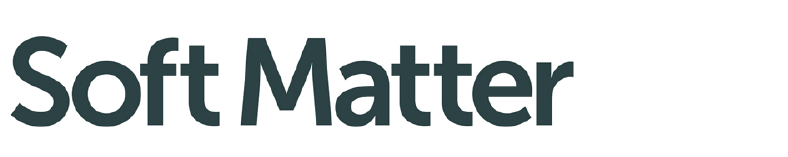}\hfill\raisebox{0pt}[0pt][0pt]{\includegraphics[height=55pt]{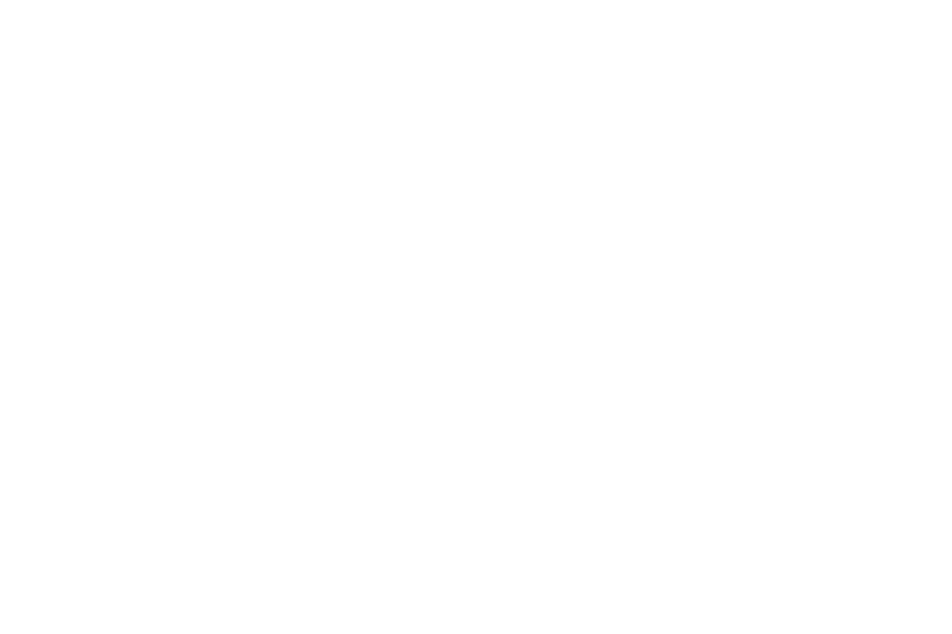}}\\[1ex]
		\includegraphics[width=18.5cm]{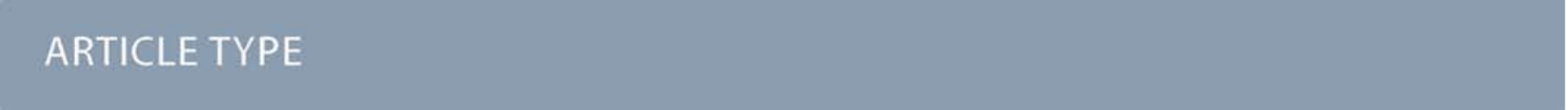}}\par
	\vspace{1em}
	\sffamily
	\begin{tabular}{m{4.5cm} p{13.5cm} }
		
		\includegraphics{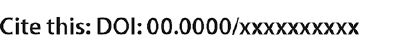} & \noindent\LARGE{\textbf{Strain localization and failure of disordered particle rafts with tunable ductility during tensile deformation}} \\
		\vspace{0.3cm} & \vspace{0.3cm} \\
		
		& \noindent\large{Hongyi Xiao,\textit{$^{a}$} Robert JS Ivancic,\textit{$^{a}$} and Douglas J Durian\textit{$^{a\ast}$}} \\
		
		\includegraphics{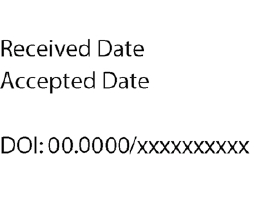} & \noindent\normalsize{
			Quasi-static tensile experiments were performed for a model disordered solid consisting of a two-dimensional raft of polydisperse floating granular particles with capillary attractions. The ductility is tuned by controlling the capillary interaction range, which varies with the particle size. During the tensile tests, after an initial period of elastic deformation, strain localization occurs and leads to the formation of a shear band at which the pillar later fails. In this process, small particles with long-ranged interactions can endure large plastic deformations without forming significant voids, while large particles with short-range interactions fail dramatically by fracturing at small deformation. Particle-level structure was measured, and the strain-localized region was found to have higher structural anisotropy than the bulk. Local interactions between anisotropic sites and particle rearrangements were the main mechanisms driving strain localization and the subsequent failure, and significant differences of such interactions exist between ductile and brittle behaviors.} 
		
	\end{tabular}

 \end{@twocolumnfalse} \vspace{0.6cm}

]

\renewcommand*\rmdefault{bch}\normalfont\upshape
\rmfamily
\section*{}
\vspace{-1cm}


\footnotetext{\textit{$^{a}$~Department of Physics and Astronomy, University of Pennsylvania, Philadelphia, PA 19104, USA } 
\textit{$^{*}$~Email:~djdurian@physics.upenn.edu} 
}


\section{Introduction}

Improving the ductility of disordered solids is an ongoing challenge as many of them have high application value but cannot withstand large plastic deformation beyond yielding, and often fail catastrophically.\cite{chen2008mechanical,greer2013shear,lan2019universal,zhang2013using} Strain localization is an important process that leads to such failures, where strain in the early stage of deformation gradually localizes into a single region that spans across the sample.\cite{manning2007strain,le2014emergence,ma2018strain} This process often results in the formation of a shear band, where the material later fails. Strain localization and shear band formation occur in a variety of disordered solids, such as metallic glasses,\cite{greer2013shear,chen2008mechanical} glassy polymers,\cite{lin2019distinguishing,ivancic2019identifying} foams,\cite{wang2006impact,katgert2008rate} and granular materials,\cite{tapia2013effect,schall2010shear,kawamoto2018all} despite the vast differences in the details of their composition. The similarity in their mechanical behavior comes from their disordered structures, which must rearrange during plastic deformation.\cite{falk1998dynamics} In a simplified picture for strain localization, early-stage local rearrangements tend to occur at sites that are structurally weak,\cite{nicolas2018deformation,manning2007strain,cubuk2017structure} which in turn further increase their susceptibility for more rearrangements. It is believed that the cooperative effects of these local rearrangements can lead to the formation of a system-spanning shear band.\cite{dansereau2019collective,karimi2018correlation,dasgupta2012microscopic} While there are several theoretical approaches that capture this process on continuum scale or mesoscale,\cite{falk1998dynamics,falk2011deformation,sollich1997rheology,dansereau2019collective} direct experimental observation of structural weakening is still lacking.

 The structural evolution during strain localization and failure for materials with different ductility has not been well described. Although the occurrence of strain localization is universal, the pathway for it and the following failure process can be different. For example, materials like wet foams, bubble rafts, and certain granular media can be highly ductile and exhibit a fluid-like behavior by sustaining large plastic deformation without forming significant voids or fractures.\cite{lauridsen2002shear,harrington2018anisotropic,kuo2012scaling,katgert2008rate} On the other hand, materials like metallic glasses and rocks can be highly brittle and form a sharp fracture after relatively little plastic deformation beyond yielding.\cite{greer2013shear,lan2019universal,zhang2013using}  Moreover, it is also possible to induce a brittle to ductile transition for a single type of material by tuning its properties such as temperature\cite{lin2019distinguishing}, particle shape,\cite{zhang2013using} internal friction,\cite{karimi2019plastic} preparation history,\cite{gu2009ductile,ozawa2018random,lin2019distinguishing} system size\cite{guo2007tensile,sopu2016brittle,cho2019crack,shi2010size}, and particle interaction.\cite{falk1999molecular,dauchot2011athermal,babu2016excess,lin2019distinguishing,karimi2019plastic} The mechanisms leading to the differences in ductility are not entirely clear, but some of these methods modify particle properties, such as friction and shape, which suggests that the transition should have a microscopic origin. Thus the interaction between local structure and local dynamics could play an important role, and this can be better understood by examining a model experimental system with tunable ductility. 

Here we focus on an approach to tune ductility used previously in simulations, which is to modify the interaction potential between particles.\cite{falk1999molecular,dauchot2011athermal,babu2016excess,lin2019distinguishing} These computational studies modified the Lennard-Jones potential in different ways, but they all showed that the ductility of the disordered material increases with increasing the characteristic interaction range between particles. While this is relatively easy to accomplish in computer simulations, controlling interaction range in an experimental system while tracking all the particles during the highly transient strain localization and failure processes is challenging. One relevant branch of experimental methods is to fabricate disordered solids by connecting particles with tunable rigid bridges\cite{schmeink2017fracture,hemmerle2016cohesive,hor2017nanoporous,jiang2018toughening} By varying the stiffness and volume of these bridges, the fracture toughness of the material measured during crack propagation can be improved.\cite{schmeink2017fracture,jiang2018toughening} Although particle-level rearrangements were not examined in these studies and the underlying mechanism for improved ductility is more related to stiffening and failure of the rigid bridges, rather than particle rearrangements, these results are certainly encouraging for designing more experimental systems with tunable particle interactions and preferably with more degrees of freedom for particles to rearrange.

Following this idea, we built and performed experiments on a model disordered solid made of a monolayer of granular particles floating at an air-oil interface (a particle raft) with capillary attractions between the particles. The capillary attraction is caused by the distortion a floating particle induces to the surrounding fluid interface. For a second particle that is nearby, this interface distortion causes an imbalance between its gravity, buoyancy, and the capillary force, which incurs a net attractive force between the two particles.\cite{nicolson1949interaction,kralchevsky2000capillary,singh2005fluid,dalbe2011aggregation,ho2019direct} In this way, the capillary attraction is often long-ranged with the characteristic interaction range being the capillary length of the liquid involved, $l_c$,\cite{nicolson1949interaction,dalbe2011aggregation} which is roughly the size of the liquid meniscus around a particle. Combined with the short-ranged repulsion between particles in contact, the interaction potential shares similarities with potentials of other particles of interest such as atoms.\cite{nicolson1949interaction} This similarity, along with the ease of observation, have made particle raft (often bubbles) a model system to demonstrate and study the physics of many crystalline and amorphous materials.\cite{bragg1947dynamical,mazuyer1989shear} Moreover, the viscous drag on the particles can be minimized by adjusting the particle velocities, so that the deformation of particle rafts can be free from basal friction,\cite{kozlowski2019dynamics} which is often a problem for using two-dimensional systems as model systems. On the other hand, the particle raft itself can be an important system in various fields such as self-assembly\cite{liu2018capillary,aubry2008micro,nishikawa2003self} and particle-coating for interfaces in applications including drug delivery and food production.\cite{f1999encapsulation,tsapis2002trojan,nishikawa2003self} These applications can benefit from better understanding of mechanical behaviors of the particle rafts\cite{bandi2011shock,mazuyer1989shear,knoche2015elasticity,cicuta2009granular,planchette2012surface,kuo2012scaling,vella2006dynamics}.

In this study, quasi-static tensile tests were performed for particle rafts and the capillary interactions were controlled by using different particle diameters, $d$, which essentially controls the characteristic interaction range in units of the particle diameter, $l_c/d$. This allowed us to observe structural changes of pillars showing different ductility, which qualitatively agrees with previous computational studies.\cite{falk1999molecular,dauchot2011athermal,babu2016excess,lin2019distinguishing} These experiments also reveal differences in the interplay between structure and dynamics during strain localization and failure for materials with different ductility.

The rest of this article is organized as follows. In Sec.~\ref{Sec2}, we describe the particle rafts, the experimental apparatus, and techniques for tracking particles and measuring the tensile force. Sec.~\ref{Sec3} demonstrates the brittle and ductile behaviors observed in the experiments. Sec.~\ref{Sec4} presents detailed analysis of structure-dynamics relations during strain localization and failure for pillars with different ductility. Sec.~\ref{Sec5} presents the conclusions.

\section{Tensile experiments of floating granular particles}
\label{Sec2}

In this study, the rafts consist of spherical particles floating at an air-oil interface. The particles are made of closed-cell Styrofoam with a density of approximately 15~kg/m$^3$. The particles are slightly polydisperse, and three batches of particles with different mean diameters $d$ were studied: $d=0.7\pm$0.1~mm, $d=1.0\pm$0.1~mm, and $d=3.3\pm$0.3~mm, as measured using a Camsizer (Retsch). The oil used in the experiments is mineral oil as in a previous study.\cite{rieser2015tunable} The surface tension is estimated to be $\gamma=27.4\pm0.7$~dyn/cm, the density is $\rho=870\pm10$~kg/m$^3$, resulting in a capillary length of $l_c=\sqrt{\gamma/\rho g}=1.8\pm0.2$~mm. The kinematic viscosity of the mineral oil is approximately $\nu=$13.5~cSt.

\begin{figure}[t]
\centering
  \includegraphics[width=8.9 cm]{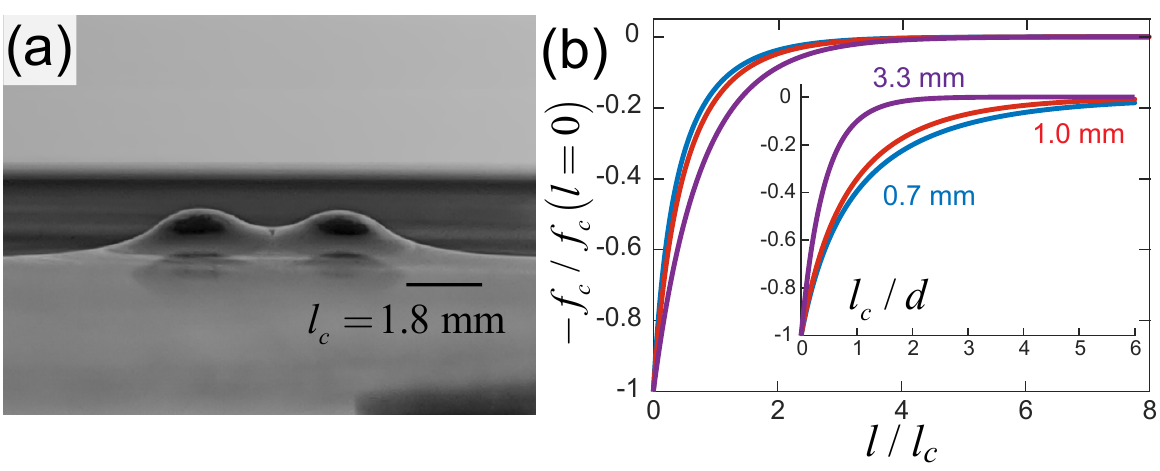}
  \caption{Demonstration of the capillary interaction: (a) Photo of two similar sized floating $d=3.3$~mm particles. (b) Normalized capillary attraction of a single pair of particles vs. separation distance normalized by the capillary length or by the average particle diameter for the three particle sizes (inset).}
  \label{fgr:capillary}
\end{figure}

An example of two floating particles is depicted in Fig.~\ref{fgr:capillary}(a). As seen, the contact angle between the particle surface and the oil is small, and the particles are pulled down by the surface tension. This type of capillary attraction was recently analyzed by Dalbe {\it et al.},\cite{dalbe2011aggregation} and the attractive force is $f_c=-CK_1[(l+d)/l_c]$. Here, $C$ is a constant depending on properties of the particles, the liquid, and wetting, $l$ is the separation distance between the two particle surfaces ($l=0$ at close contact), and $K_1(X)$ represents the modified Bessel function of the second kind and first order. Figure~\ref{fgr:capillary}(b) shows how $f_c$ (normalized by $f_c$ at $l=0$) decays with $l$ (normalized by $l_c$). Here, the attraction decays rapidly over $l_c$ and becomes negligible after $2l_c$. By normalizing $l$ using $d$, the inset in Fig.~\ref{fgr:capillary}(b) shows that $f_c$ for the smaller particles (1.0~mm and 0.7~mm) can extend over a few $d$, while for the 3.3~mm particles $f_c$ decays rapidly within one $d$. Thus, in units of $d$, the range of the capillary interaction increases with decreased particle diameter. Although the liquid surface distortion becomes much more complicated in a dense packing, resulting in many-body contributions to the potential energy, this difference in the interaction range should be preserved, at least over voids when a pair of particle are not completely blocked by other particles.\cite{nishikawa2003self}

\begin{figure}[h]
\centering
  \includegraphics[width=8.5 cm]{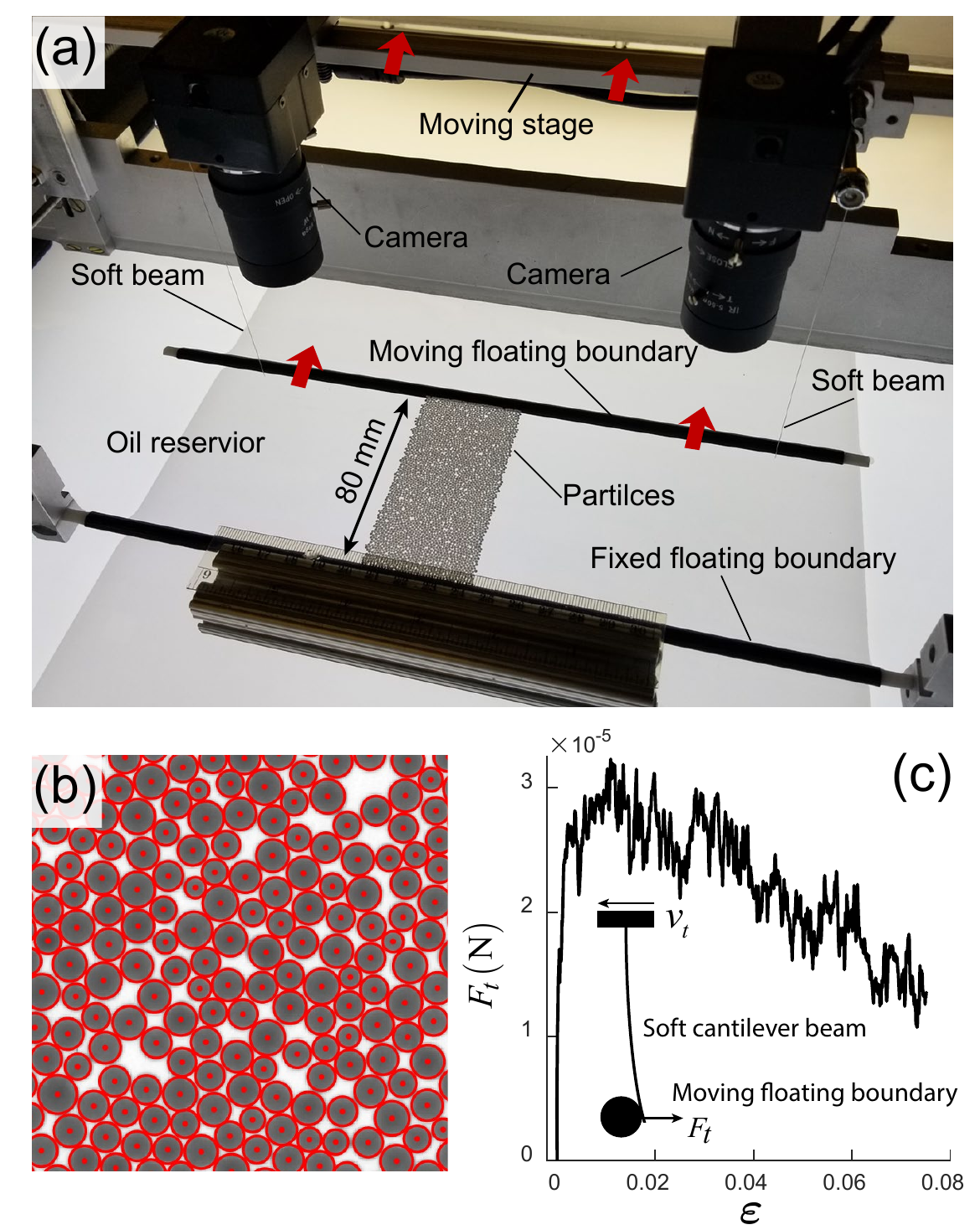}
  \caption{The experimental apparatus. (a) A photo of the experimental apparatus. (b) An example of particle tracking with the tracked centers marked by dots and the tracked radii displayed using circles. (c) An example of measured tensile force vs. global strain for a single experiment with 1~mm particles.}
  \label{fgr:setup}
\end{figure}

The experimental apparatus used here is based on a previous setup designed to study plastic deformation of granular materials.\cite{harrington2018anisotropic,harrington2019machine,rieser2015deformation,rieser2015tunable,cubuk2017structure,harrington2020stagnant} It is capable of applying a well-controlled uniaxial global strain to a two-dimensional granular material while tracking all of the particle positions and measuring the global resistance force. In this study, this apparatus was adapted to performing tensile tests for the particle rafts, which is shown in Fig.~\ref{fgr:setup}(a). Here, a pillar of particles floating on the oil is sandwiched by two boundaries made of hollow carbon fiber tubes that also float on their own. The the boundary on the bottom (of the picture) is fixed while the top boundary is driven away from the bottom boundary by a stage moving\cite{rieser2015deformation,harrington2018anisotropic} at a constant tensile velocity $v_t$. The tensile velocity is conveyed to the floating top boundary to apply a global tensile strain to the pillar via two soft cantilever beams that connect the moving stage and the top boundary (to be described later). For the 0.7~mm and 1.0~mm particles, the particles are naturally attracted and locked to the boundaries by the capillary attraction, which is stronger than the particle-particle attraction. For the 3.3~mm particles, the particle-boundary attraction is not as strong, so a layer of particles were glued to the boundary to prevent boundary detachment. In both cases, no relative motion between the particles and the boundaries were found during the experiments.

For studying the structure of the particle rafts it is important to prepare a well-shaped rectangular pillar made of a strictly single layer of densely packed particles. To meet these requirements, we first initiated a thin pillar (less than 5$d$ wide) that connected the two boundaries, and then we grew the pillar by dropping particles near its two sides, and let the particles assemble to the existing pillar driven by the capillary attraction, until the pillar reaches the desired shape. In this way, we can obtain dense disordered packing with no particle overlaps or large voids, see Fig.~\ref{fgr:setup}(a) and (b), and the fluctuation of the boundary shape is typically smaller than 1$d$. Previous simulations of small-scale tensile tests suggest that the occurrence of strain localization and the formation of the shear band is not sensitive to the system size as long as it is larger than 3-4 times of the shear band size.\cite{shi2010size} Here, we also found that shear band formation is not sensitive to the system size when the pillar height is larger than approximately 40$d$. We also varied the height/width ratio from 1:1 to 4:1 and the phenomenon is also not sensitive to it. Based on these observations, we selected a reasonable pillar size by setting the pillar height to be $L_0=80d$ and width to be $W_0=40d$ for all the particle sizes. Given the selected system size, we chose a tensile strain rate of $\dot{\epsilon}=v_t/L_0=$1.3$\times10^{-5}$~s$^{-1}$ for all the particle sizes, corresponding to $v_t=1.04\times10^{-3}$~$d$/s. This results in a capillary number of $Ca=\mu v_ t/\gamma$ that is on the order of $10^{-6}$, where $\mu=\nu \rho$, and a Reynolds number of $Re=v_td/\nu$ that is on the order of $10^{-4}$. This suggests that hydrodynamic forces and viscous forces are much smaller than the capillary attractions in the system, and the experiments were in a quasi-static regime.

For each experiment, images with a resolution of 2048$\times$2048~px$^2$ were recorded by a JAI/Pulnix TM-4200CL camera with a time interval of 0.75~s, corresponding to a displacement of the moving boundary of 7.8$\times 10^{-4}d$. The positions and radii of all the particles were tracked using a previously developed algorithm with a sub-pixel accuracy,\cite{rieser2015deformation,harrington2018anisotropic} and examples of the tracked particle center and diameter are plotted on top of a raw experimental image in Fig.~\ref{fgr:setup}(b). To reduce noise, we further applied a Gaussian filter to the measured positions with a time window corresponding to a moving boundary displacement of approximately 1/15$d$, similar to our previous work.\cite{harrington2018anisotropic,harrington2018anisotropic} And then particle velocities, $\mathbf{v}$, were calculated based on the filtered positions.

The global tensile force was also measured during the experiment. The tensile force, $F_t$, in these experiments ranges from $10^{-6}$ to $10^{-3}$~N, which is too small for typical commercial force sensors. Here the force was determined by measuring the deflection that the soft cantilever beams generated while pushing on the moving floating boundary, as depicted in Fig.~\ref{fgr:setup}(c). The deflection has a linear relationship with the tensile force,\cite{he2015measured} and it was measured as the relative displacement between the moving boundary and the moving stage, which was monitored using two industrial webcams mounted on the stage, see Fig.~\ref{fgr:setup}(a). The beams are made of stainless steel, and the stiffness of the beams (controlled by diameter and length) were chosen so that observable deflections can be generated in the experiments. The maximum deflection was controlled to be slightly below 1~mm, with the resolution of the webcams being approximately 90~px/mm. In the experiments, the deflection was measured with a time interval of 5.8~s, and a moving average with a windows size of four frames was applied. The beam stiffness was separately calibrated using a commercial 10~g force sensor (Transducer Techniques). As demonstrated in Fig.~\ref{fgr:setup}(c), this force measurement procedure can clearly capture the general trend of $F_t$ in the experiment including its sharp initial increase. The major source of error comes from the fact that the actual length of the beam (between the two contact points) could be slightly different between the calibration and the experiment, which could induce an error in the conversion from deflection to force that is within 5\%. However, the shape of the data is not affected by this.

For each particle size, 50 tensile deformation tests were performed to achieve good statistics on local deformation and structural changes, as discussed in the following sections.

\section{Observations of brittle and ductile behaviors}
\label{Sec3}

In the tensile experiments, a transition from brittle to ductile behavior can be clearly observed as the particle size decreases. To better visualize the differences, we quantify the local deviatoric strain rate, $J_2$, in the deforming pillar at different global strains. The calculation of $J_2$ is detailed in our previous studies,\cite{harrington2018anisotropic,harrington2020stagnant} which starts from a Delaunay triangulation of instantaneous particle positions. For a single triangle, we calculated a local strain rate tensor $\dot{\mathbf{e}}$ based on the velocity $\mathbf{v}$ of the particles on its vertices using the constant strain triangle formalism,~\cite{Cook1974}

\begin{equation}
	\left(\begin{array}{c}
		v_x(x,y)-v_{x,c} \\
		v_y(x,y)-v_{y,c}
	\end{array}\right)
	=
	\left(\begin{array}{cc}
		\dot{e}_{11} & \dot{e}_{12} \\
		\dot{e}_{21} & \dot{e}_{22}
	\end{array}\right)
	\left(\begin{array}{c}
		x \\
		y
	\end{array}\right),
\end{equation}
where $x$ and $y$ are Cartesian coordinates relative to the triangle centroid, and $v_{x,c}$ and $v_{y,c}$ are the velocity at the centroid (to be computed). From the symmetric portion $\dot{\mathbf{\epsilon}}=(\dot{e}_{ij}+\dot{e}_{ji})/2$ we calculate the local deviatoric strain rate $J_2$,
\begin{equation}
\label{eq:J2}
	J_2 = \frac{1}{2}\sqrt{(\dot{\epsilon}_{11}-\dot{\epsilon}_{22})^2+4\dot{\epsilon}_{12}^2}.
\end{equation}
Following our previous work,\cite{harrington2018anisotropic,harrington2020stagnant} we normalize $J_2$ by a characteristic strain rate $v_t/d$ during the analysis.\cite{harrington2018anisotropic}

\begin{figure}[t]
\centering
  \includegraphics[width=9.2 cm]{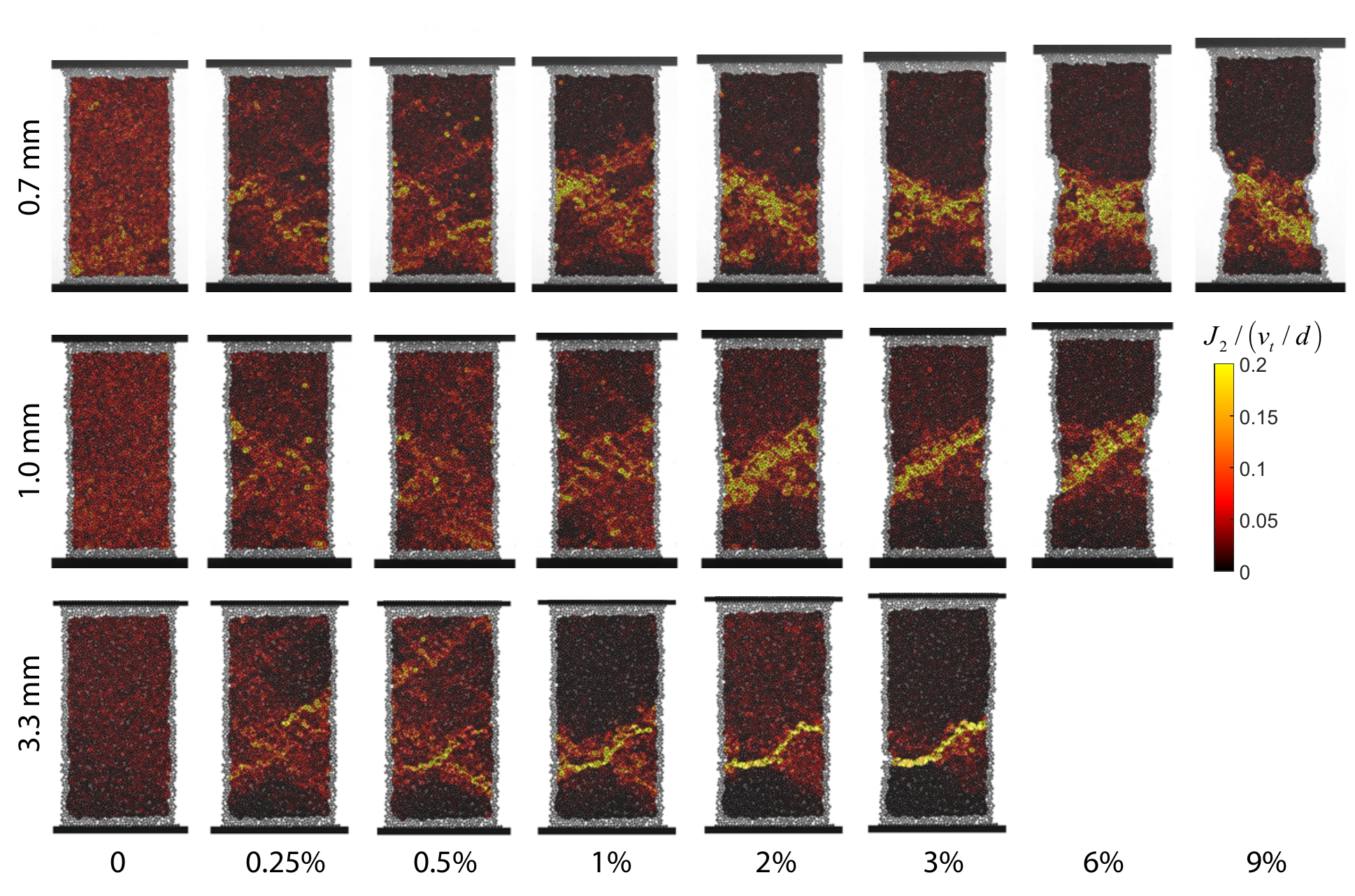}
  \caption{Normalized $J_2$ for examples of different particle sizes at different global strains. In each pillar, local $J_2$ for each triangle is plotted on top of the original experimental image. }
  \label{fgr:j2map}
\end{figure}

While $J_2$ is good for identifying deviatoric deformation, we also calculated a second local quantity, $D^2_{min}$, which specifically picks out the non-affine part of the deformation that corresponds to local particle rearrangements.\cite{falk1998dynamics} The calculation of $D^2_{min}$ is based on the change of particle positions, $\mathbf{r}_i$, between two consecutive frames with a time interval of $\Delta t$. For each particle $i$, a best-fit local affine deformation matrix, $\mathbf{E}$, can be computed, and then the non-affine displacement, $D^2_{min}$, associated with particle $i$ can be calculated,\cite{li2015deformation,falk2011deformation,harrington2019machine}
\begin{equation}
\label{eq:D2}
D^2_{min,i}(t,\Delta t) = \frac{1}{n}\sum_{i=1}^{n}\left|\mathbf{r}_{ji}(t+\Delta t)-\mathbf{E}\mathbf{{r}}_{ji}(t)\right|^2,
\end{equation}
where $\mathbf{{r}}_{ji}=\mathbf{r}_j-\mathbf{r}_i$ is the relative position between particle $i$ and its neighbor $j$. Here we selected a time interval that corresponds to a global tensile strain of 0.2\%, which is a typical duration for a rearrangement. The search radius for neighbors is set to be 1.25$d$ so that the first shell of neighbors is included, which is roughly the same group of particles that share triangles with the center particle. In the following analysis, we normalize $D^2_{min}$ by $d^2$.\cite{harrington2019machine,li2015deformation}

\begin{figure}[t]
\centering
  \includegraphics[width=9.2 cm]{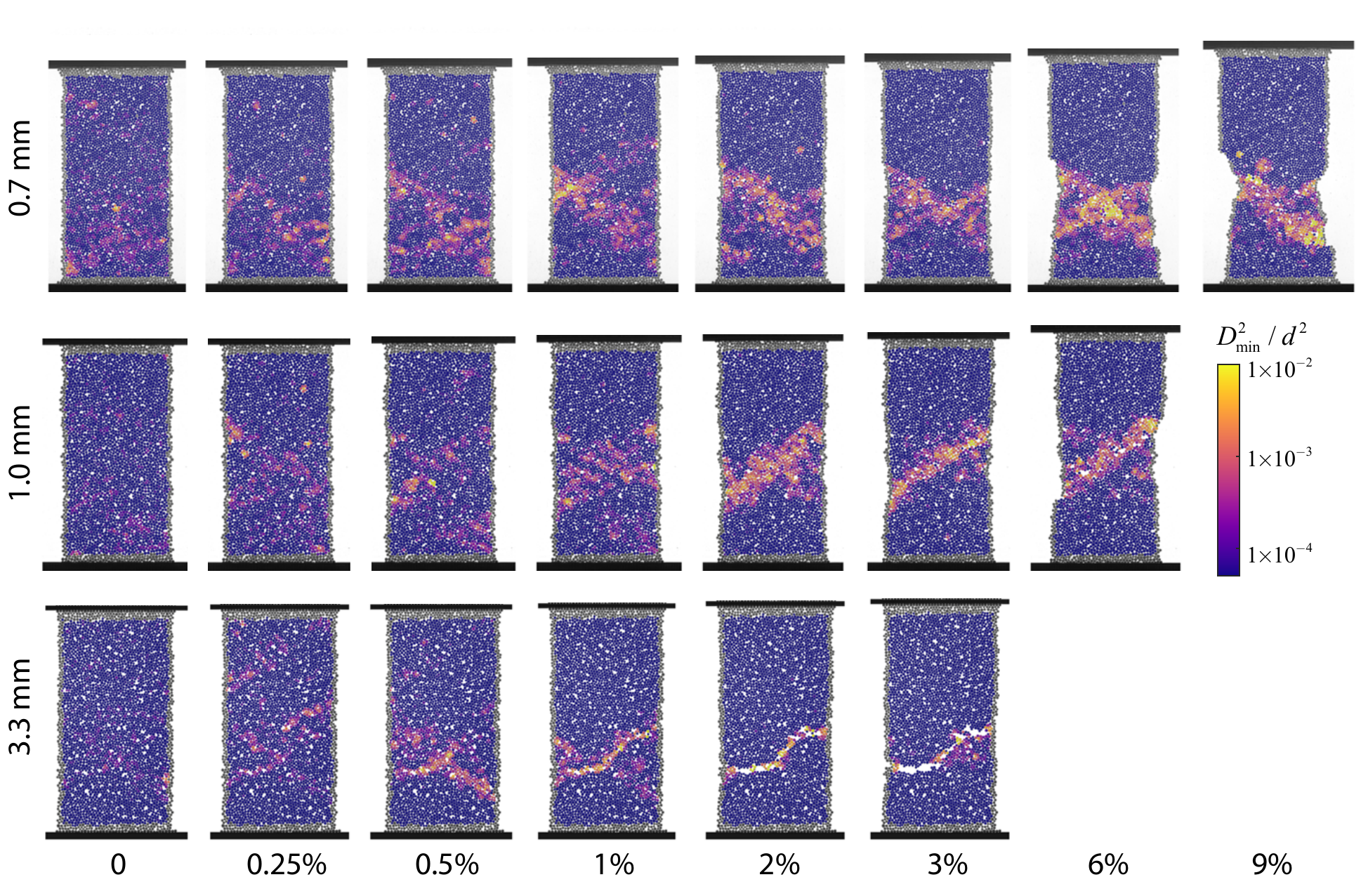}
  \caption{Normalized $D^2_{min}$ for examples of different particle sizes at different global strains (same cases as Fig.\ref{fgr:j2map}). In each pillar, the particles are colored by their $D^2_{min}$ values.}
  \label{fgr:d2map}
\end{figure}

Using $J_2$ and $D^{2}_{min}$, we demonstrate examples of strain localization and failure for the three particle sizes at different global strains, $\epsilon=(L-L_0)/L_0$, where $L$ is the instantaneous pillar height. These results are shown in Figs.~\ref{fgr:j2map} and~\ref{fgr:d2map}, where triangles colored by $J_2$ or particles colored by $D^2_{min}$ are plotted on top of experimental images. Note that the particles near the boundaries are left out of the analysis.

Figures~\ref{fgr:j2map} and~\ref{fgr:d2map} show that while strain localization and failure occur for all particle sizes, qualitative differences are found between them. At the very beginning of the experiments ($\epsilon\approx0$), the local $J_2$ is uniformly distributed throughout the pillar for all three particle sizes, and little significant rearrangement occurs, indicating an initial elastic-like deformation. The magnitude of $J_2$ appears to slightly decrease with increased particle size. As $\epsilon$ proceeds to 0.25\% and 0.5\%, $J_2$ is still fairly spread out in the pillar, but its distribution is clearly non-uniform. This hint of strain localization is accompanied by the appearance of some high $D^2_{min}$ values, indicating some small-scale particle rearrangements, at locations where $J_2$ is also higher. The difference for the three particle sizes is small at this stage, but starts to show up as $\epsilon$ increases to around 1\%. Now, the deformation for the 3.3~mm case is strongly localized to a single region, evident by a narrow and system-spanning band with high $J_2$. However, for the 0.7~mm and 1.0~mm cases, $J_2$ is distributed in a relatively wider region, and $J_2$ gradually concentrates into a system-spanning band as $\epsilon$ approaches 2\%. In the meantime, $D^2_{min}$ in the bands for the smaller particles is higher and has a wider spread comparing to that of the 3.3~mm particles, indicating that the smaller particles are more capable of rearranging themselves to accommodate the global deformation. 

The emergence of the strain-localized region can be treated as the onset of failure. As the global strain further increases, a significant difference in the ductility of the pillars can be observed. For the 3.3~mm particles, a fracture develops at $\epsilon=2\%$ and $3\%$ from the strain-localized region. At this point, the high $J_2$ at the fracture is mainly due to the growth of voids, which can be clearly seen in the $D^2_{min}$ images as the white space between particles. Only small rearrangements exist around the fracture, and the pillar breaks apart with little shape deformation, showing typical a brittle behavior. On the other hand, pillars formed from smaller particles are more ductile. For 0.7~mm particles, particle rearrangements occur over extended region of the sample ($\epsilon=3-9\%$), and no significant system-spanning fracture exists. Instead, a long-lasting shear occurs between the upper half and lower half of the strain-localized region, which is a shear band. The overall shape of the pillar deforms significantly and the shear band region continues to become narrower as $\epsilon$ increases, showing a typical ductile behavior. The failure of 1.0~mm particles is close to that of 0.7~mm particles, but by examining all 50 runs, we found there is typically a few voids growing at large $\epsilon$, similar to the 3.3~mm particles.

Throughout the 50 tensile tests performed for each particle size, strain localization and failure are consistent with the examples in Figs.~\ref{fgr:j2map} and~\ref{fgr:d2map}. For each particle size, the shear band appears at similar global strains with similar characteristics, while its vertical location can vary throughout the height of the pillar. The average angle of the shear band with the horizontal direction is 26.3$\pm5.2^\circ$ for $d=0.7$~mm, 27.5$\pm4.3^\circ$ for $d=1.0$~mm, and 26.6$\pm6.9^\circ$ for $d=3.3$~mm. Note that for the 3.3~mm particles, the failure planes still have the same inclination as that for the smaller particles, which could arise from similarities in their early-stage strain localization process, so we also refer to them as shear bands. The value of this inclination is different from the 45$^\circ$ that is commonly seen in materials like metallic glasses and polymers.\cite{greer2013shear,chen2008mechanical,lin2019distinguishing,ivancic2019identifying} This difference is possibly a result of particle friction that is unique to granular materials.\cite{le2014biaxial,le2014emergence} This could be caused by a combined effect of a local Mohr-Coulomb failure criterion and long-range elastic interactions between the failure sites (local rearrangements).\cite{karimi2018correlation}

\begin{figure}[t]
\centering
  \includegraphics[width= 8.9 cm]{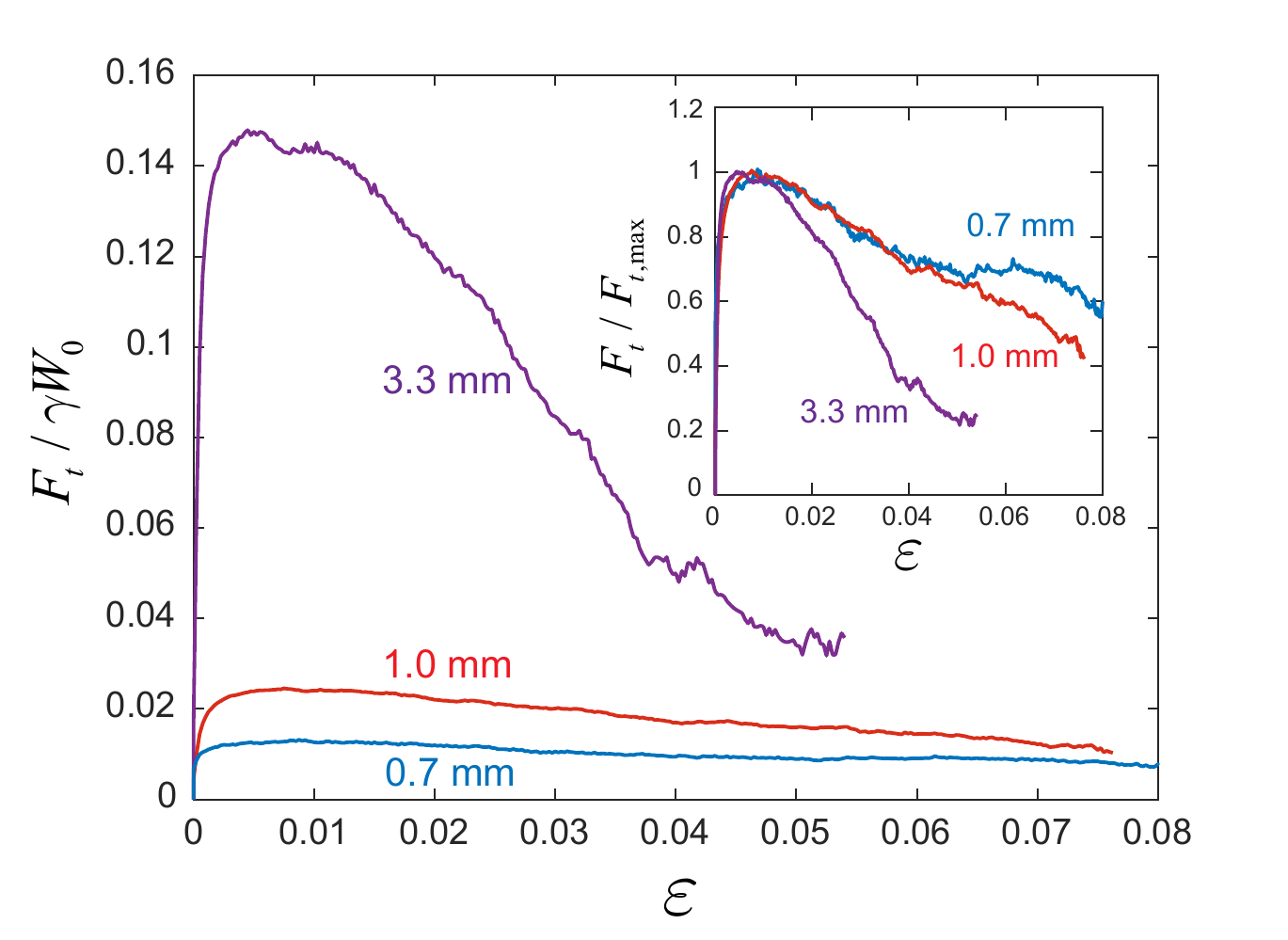}
  \caption{Normalized stress-strain curve averaged over 50 experiments for each particle size. Inset: tensile force normalized by the maximum measured tensile force, $F_{t,max}$.}
  \label{fgr:force}
\end{figure}

Another way to examine the brittle-to-ductile transition is to study the stress-strain curves for pillars of different particle sizes. Here, a normalized global tensile stress is calculated by normalizing the measured tensile stress, $F_t/W_0d$, with a characteristic Laplace pressure $\gamma/d$, giving $F_t/\gamma W_0$. The result of this straightforward normalization is shown in Fig.~\ref{fgr:force}, with each curve calculated as an average of the 50 experiments performed for each particle size. The maximum stress (sometimes called the ultimate tensile strength) increases with $d$, with the maximum stress of 3.3~mm particles being about an order of magnitude larger than that for the 0.7~mm particles. This is possibly due to the fact that larger particles can bring larger distortion to the liquid surface and thus have larger capillary attractions.\cite{dalbe2011aggregation} The stress difference is comparable to the difference in their Bond number, $Bo=d^2/4l^2_c$, which compares gravitational forces with the capillary forces. This could be a useful reference for future study on the capillary attraction in a dense packing.

Here, we mainly utilize the stress-strain curves to understand different periods in the tensile deformation. Initially, an elastic behavior is observed for all $d$ with the tensile stress rapidly increases. This is the period where $J_2$ is uniform and little rearrangements occur (Figs.~\ref{fgr:j2map} and~\ref{fgr:d2map}). For $\epsilon=$0.25\%-1\%, the increase of the tensile stress slows down until it flattens for the 0.7~mm and 1.0~mm particles. For the 3.3~mm particles, the stress reaches a peak at $\epsilon=$0.6\%, and then slightly drops before it flattens. This could be a stress overshoot at yielding that is more significant for brittle materials\cite{ozawa2018random}, but it could also be a result of lack of averaging and needs further investigation with larger systems and better averaging. The slowing down of the stress increase coincides well with the starting of strain localization and the appearance of plastic rearrangements, and this period can be considered as plastic deformation beyond yielding. The third period is the failure process evident by the decay of stress. To show it more clearly, we normalize the mean tensile stress by its maximum, which gives $F_t/F_{t,max}$, see the inset of Fig.~\ref{fgr:force}. This normalization shows a clear trend that the rate of decay is slower for smaller particles, which further confirms their higher ductility. The relatively rapid decay of strength for the 3.3~mm particles is mainly due to the growth and merging of voids, while the slower decay of strength for the 0.7~m is mainly due to the decrease of cross-sectional area. The strength decay of the 1.0~mm particles is intermediate in a way that it follows the curve of the more ductile 0.7~mm particles until approximately $\epsilon=5\%$, and falls off, which could be an effect of void growth. 

These results show that we experimentally achieved a brittle-to-ductile transition by decreasing $d$, which corresponds to increasing the interaction range of the capillary attraction. This transition agrees qualitatively with transitions found in previous numerical simulations that modified Lennard-Jones-like potentials.\cite{falk1999molecular,dauchot2011athermal,babu2016excess,lin2019distinguishing} 

\section{Structural changes during strain localization and failure}
\label{Sec4}

\subsection{Quantifying structural changes}

In this section, we quantify structural changes during tensile deformation by examining the local structural anisotropy and study how it interacts with local deformation. The structural anisotropy can be quantified using different approaches, such as free volume,~\cite{Turnbull1961} Voronoi cell size and shape,~\cite{SchroderTurkEPL2010,MorseCorwinPRL2014} local topology,\cite{cao2018structural} and machine learning.~\cite{cubuk2015identifying,cubuk2017structure} For this study, we focus on a quantity that, like $J_2$, is defined over Delaunay triangles: The area-weighted divergence of the particle center-to-Voronoi cell centroid vector field, $Q_k$,~\cite{RieserPRL2016} 
\begin{equation}
\label{eq:Qk}
	Q_k = \nabla \cdot \mathbf{C}_k \frac{A_k}{\langle A \rangle}.
\end{equation}
For a Delaunay triangle $k$, $C_k$ is the vector field pointing from particle centers to the centroid of corresponding Voronoi cells, $A_k$ is the area of the triangle, and $\langle A \rangle$ is the average area of all triangles. By construction, the average $Q_k$ over an entire packing is zero. Positive values of $Q_k$ tend to correspond to closely packed, or ``overpacked,'' sites, while negative values correspond to voids, or ``underpacked,'' sites.  The distribution of $Q_k$ value was previously observed to be nearly Gaussian except for a noticeable tail of underpacked regions.\cite{RieserPRL2016}

\begin{figure}[t]
\centering
  \includegraphics[width= 8.9 cm]{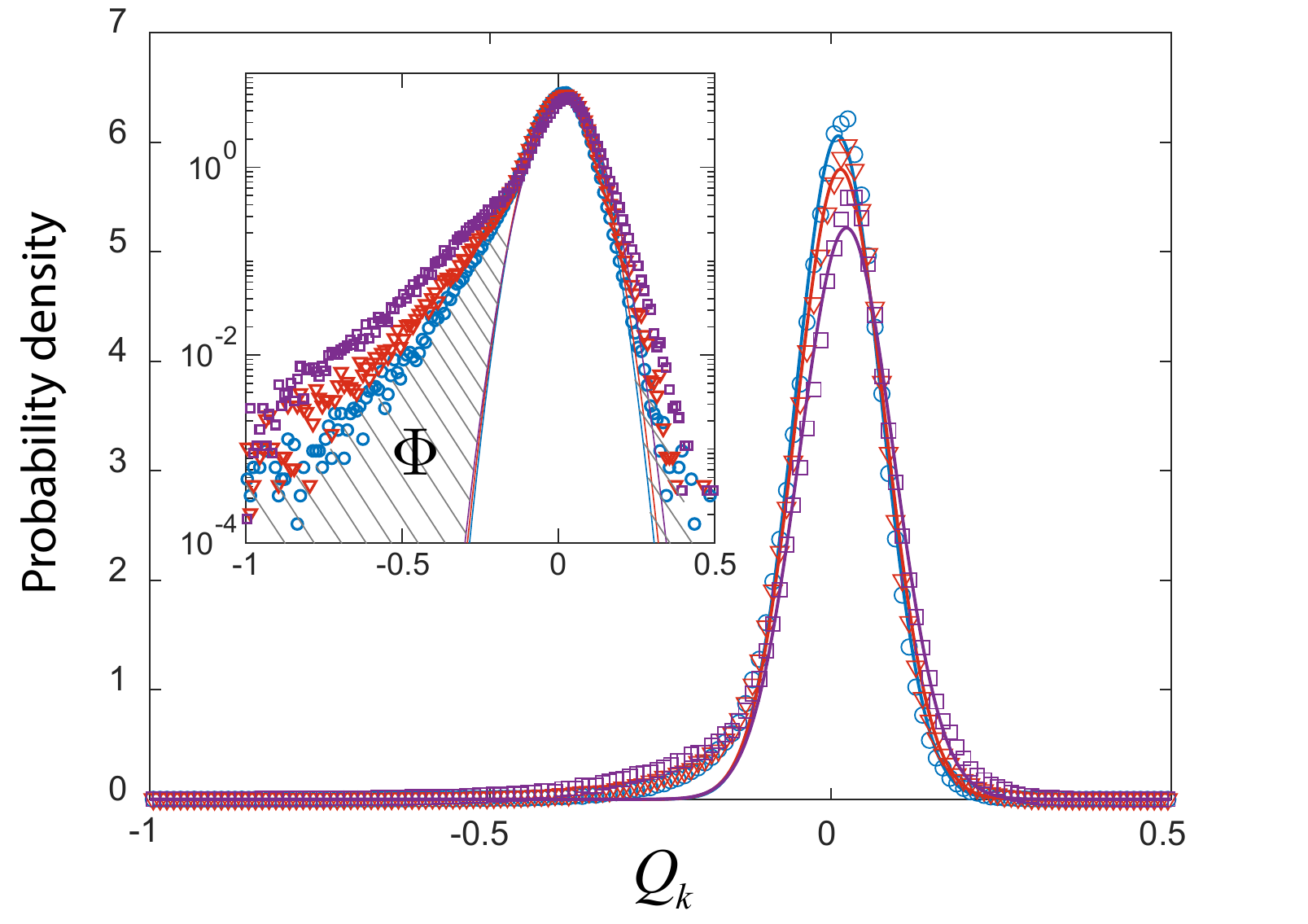}
  \caption{Initial $Q_K$ distribution calculated using 50 experiments for $d=0.7$~mm (blue circles), $d=1.0$~mm (orange triangles), and $d=3.3$~mm (purple squares). Inset shows the same plot with the vertical axis on log-scale. The curves are Gaussian fits to the region $-0.15<Q_k<0.15$ for the three sizes and are colored accordingly.}
  \label{fgr:qk_dist}
\end{figure}

Figure~\ref{fgr:qk_dist} shows the initial $Q_k$ distribution calculated using initial particle positions in all 50 experiments for each particle size. The majority of $Q_k$ resides in the region around zero with a Gaussian-like distribution,\cite{RieserPRL2016,harrington2018anisotropic,harrington2020stagnant} which is made clear by plotting a Gaussian fit calculated using $-0.15<Q_k<0.15$ for each particle size. For $Q_k<-0.15$, the distribution deviates from Gaussian and becomes exponential-like instead, see the inset of Fig.~\ref{fgr:qk_dist}. This exponential tail corresponds to the existence of highly underpacked sites distributed in the pillar. For the three particle sizes, a difference in this tail exists (inset of Fig.~\ref{fgr:qk_dist}), where the decay of the probability density as $Q_k$ decreases is slower for larger particles, meaning that the portion of highly underpacked sites is larger for more brittle materials. The tails in the $Q_k$ distributions for $Q_k>0.15$ also deviate from the Gaussian fits following a similar trend with differences between the three particle sizes. Note that these pillars are prepared following the same procedure and the initial packing fraction is similar for the three sizes, which is 0.77$\pm$0.01 for 0.7~mm, 0.78$\pm$0.01 for 1.0~mm, and 0.77$\pm$0.01 for 3.3~mm (averaged over 50 experiments). The difference in the shape of the tails should come from the way particles assemble into the pillar during the preparation, which is dictated by their capillary attraction. 

\begin{figure}[t]
\centering
  \includegraphics[width= 8.9 cm]{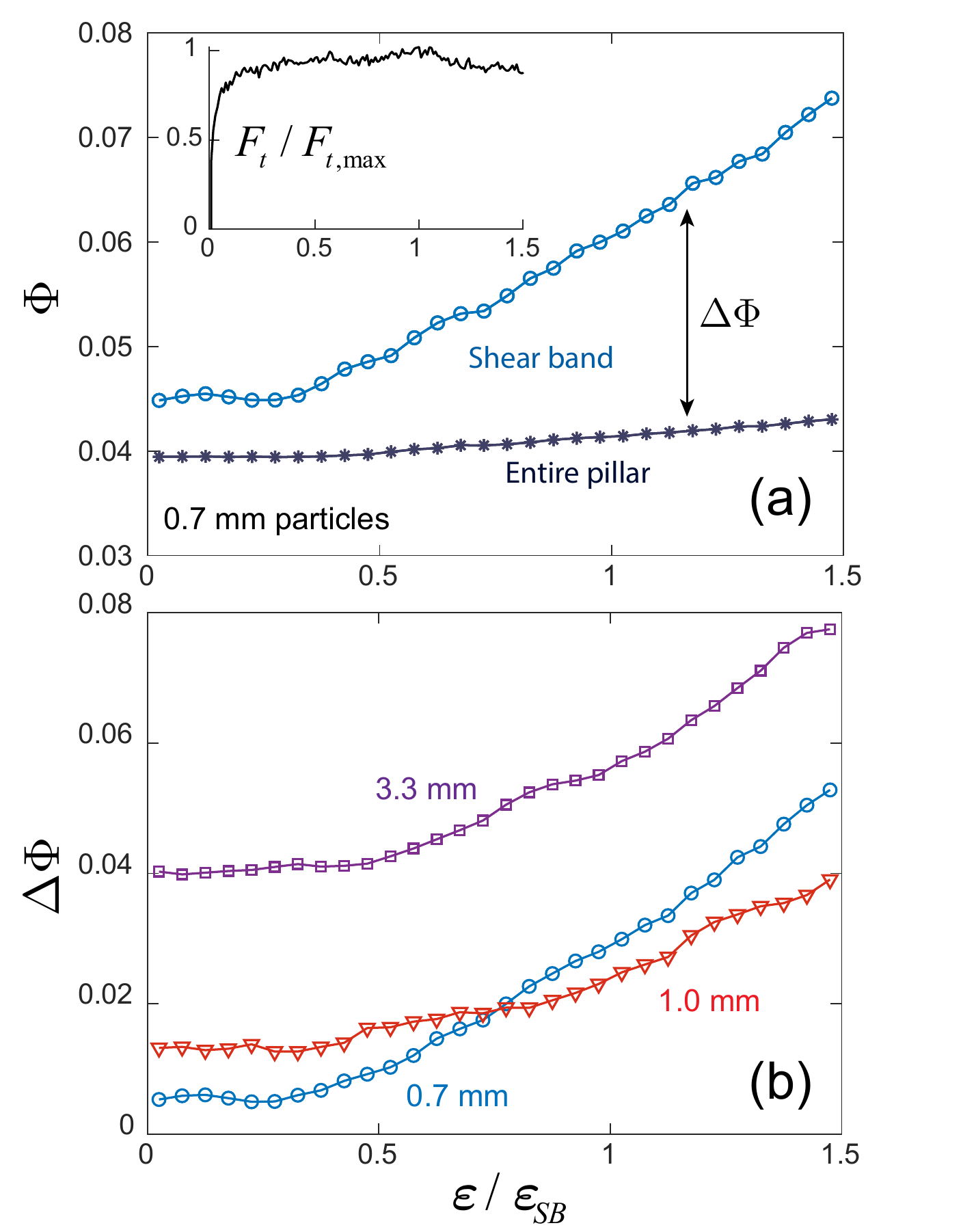}
  \caption{Excess portion $\Phi$ calculated from the $Q_k$ distribution. (a) $\Phi$ vs. $\epsilon / \epsilon_{SB}$ calculated for the entire pillar (dark asterisks) and for the shear band region only (blue circles), for $d=0.7$~mm. Inset shows averaged $F_t/F_{t,max}$ vs. $\epsilon / \epsilon_{SB}$. (b) $\Delta \Phi$ vs. $\epsilon / \epsilon_{SB}$ calculated for $d=0.7$~mm (blue circles), $d=1.0$~mm (orange triangles), and $d=3.3$~mm (purple squares). }
  \label{fgr:qk_phi}
\end{figure}

Previous studies suggest that the distribution of $Q_k$ is related to important dynamic processes in disordered solids including jamming transition,\cite{RieserPRL2016} structural strength,\cite{harrington2018anisotropic} and shear band formation.\cite{harrington2020stagnant} To study the significance of structure during strain localization, we first quantify the tails in the distribution by measuring the area difference between the probability density of $Q_k$ distribution, $p(Q_k)$, and the corresponding Gaussian fit, $p_G(Q_k)$, which is the shaded area in the inset of Fig.~\ref{fgr:qk_dist}, in a linear scale. We refer to it as the excess area, $\Phi$, and it is calculated as
\begin{equation}
\label{eq:Phi}
\Phi = \int_{-\infty}^{Q^{-}_{k}}\left(p(Q_k)-p_G(Q_k)\right) dQ_k + \int_{Q^{+}_{k}}^{\infty}\left(p(Q_k)-p_G(Q_k)\right) dQ_k,
\end{equation}
where the integration limits are, $Q^{-}_{k}=-0.15$ and $Q^{+}_{k}=0.15$ for all $d$, which correspond to where $p(Q_k)$ starts deviating from $p_G(Q_k)$. In Fig.~\ref{fgr:qk_phi}, we study how $\Phi$ changes as the global tensile strain increases. Here, we first normalize the global strain, $\epsilon$, by the strain when the shear band appears, $\epsilon_{SB}$, which is set to be the strain when the measured tensile force starts to decay. The average $\epsilon_{SB}$ over 50 experiments is 1.3$\pm0.4\%$ for 0.7~mm particles, 1.2$\pm0.4\%$ for 1.0~mm particles, and 1.0$\pm0.3\%$ for 3.3~mm particles. Figure~\ref{fgr:qk_phi}(a) shows $\Phi$ vs.\ $\epsilon/\epsilon_{SB}$ averaged over 50 experiments for 0.7~mm particles as an example. To compare the structural differences between the shear band region and the bulk, we examine $\Phi$ calculated using only triangles in the region that would develop into a shear band (assuming a thickness of 3$d$), and compare it to $\Phi$ for the bulk calculated using all triangles in the pillar. Note that these results are not sensitive to either the specific choice of the integration limits in Eq.~\ref{eq:Phi} or to the thickness of the shear band region. In addition, we also show the change of the normalized tensile stress, $F_t/F_{t,max}$, vs.\ $\epsilon/\epsilon_{SB}$ in the inset of Fig.~\ref{fgr:qk_phi}(a). 

The comparison between $\Phi$ for the shear band region and the bulk shows that $\Phi$ in the shear band region is initially higher, indicating that strain localization favors locations that have higher packing anisotropy. In the elastic period ($\epsilon/\epsilon_{SB}$<0.3), $\Phi$ remains relatively unchanged, indicating little structural change during elastic deformation when the tensile stress quickly builds up. For $\epsilon/\epsilon_{SB}>0.3$, as the pillar enters the plastic regime with initiation of rearrangements and strain localization, $\Phi$ in the shear band region starts to increase at a rate that is much faster than the rate for the bulk. The increase of packing anisotropy in the shear band region coincides well with the appearance of local rearrangements, suggesting a strong structure-dynamics coupling.  No significant transition of $\Phi$ is found at $\epsilon/\epsilon_{SB}=1$, indicating that the structure change during the initiation of the shear band is rather smooth.

Figure~\ref{fgr:qk_phi}(b) shows the difference in $\Phi$ between the shear band region and the bulk, $\Delta \Phi$, as a function of $\epsilon/\epsilon_{SB}$. The initially higher structural anisotropy in the shear band region is consistent for all three particle sizes, evident by the positive $\Delta \Phi$, and it also increases with the particle size. For all three sizes, $\Delta\Phi$ also experiences a relatively unchanged interval during elastic deformation, before it starts to increase at a global strain that coincides well with strain localization. {\it Thus, there is a strong structural signal in the strain localization process that exists long before the shear band actually appears, and it is magnified by plastic rearrangements in the early stage of the deformation.} The higher initial structural anisotropy in the shear band region is reminiscent of our previous computational study of pulled polymer nanopillars where the location of the shear band can be predicted with high accuracy using the initial structural information,\cite{ivancic2019identifying} indicating that structural difference is a universal factor that drives strain localization of disordered solids. The implications of this structural difference in the $Q_k$ distribution will be interpreted by particle-level relations between rearrangements and local deformation in the following subsection. 

\subsection{Structure-dynamic relations during strain localization}

\begin{figure}[t]
\centering
  \includegraphics[width= 8.35 cm]{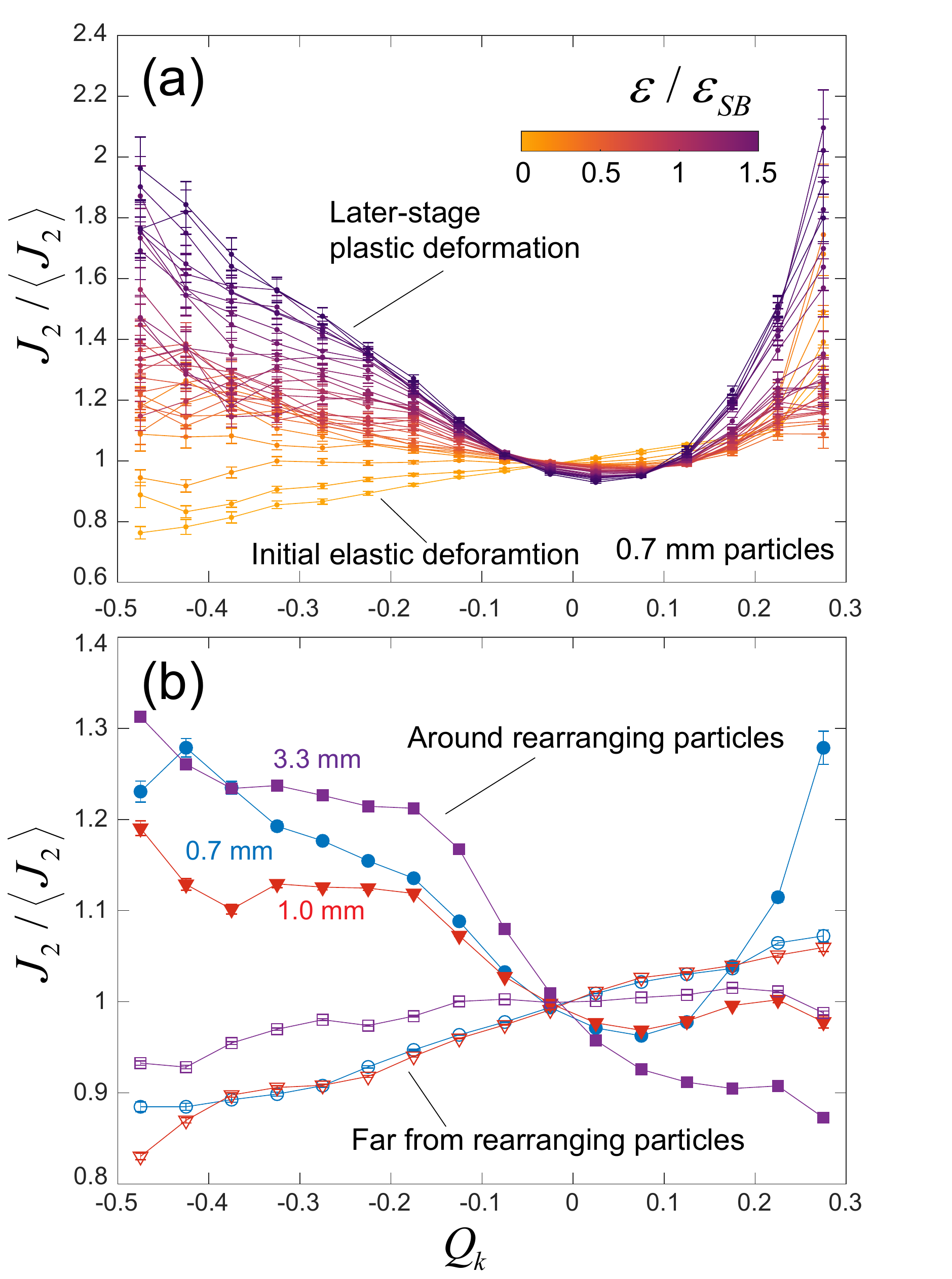}
  \caption{Relations between structure and local deviatoric deformation. (a) Triangle $J_2$ averaged by binning according to their $Q_k$ values for $d=0.7$~mm particles at different global strains. All triangles in the pillar are used and $\left<J_2\right>$ is their average at each global strain. (b) $J_2/\left<J_2\right>$ vs. $Q_k$ calculated for triangles around rearranging particles (closed) and far from rearranging particles (open) for for $d=0.7$~mm (blue circles), $d=1.0$~mm (orange triangles), and $d=3.3$~mm (purple squares). Error bars represent standard errors in both plots.}
  \label{fgr:j2_qk}
\end{figure}

One straightforward way to examine local relations between structural anisotropy and local deformation is to bin the deviatoric strain rate, $J_2$, of a triangle according to its $Q_k$ value, and average over all triangles within the bins.\cite{harrington2018anisotropic,harrington2020stagnant} Results of the bin-averaged $J_2$--$Q_k$ relation during plastic deformation of disordered solids often show a ``V-shape'', where $J_2$ is high at highly positive and highly negative $Q_k$, and $J_2$ is low around $Q_k=0$.\cite{harrington2018anisotropic,harrington2020stagnant} Here, we examine the $J_2$--$Q_k$ relation for the tensile experiments. Different from our previous experiments with a frictional substrate leading to highly localized plastic deformations,\cite{harrington2020stagnant,harrington2018anisotropic} the particle raft experiments show a well defined elastic regime where the deformation is delocalized and no plastic rearrangement occurs, which could lead to a different relation. This motivated us to compute the $J_2$--$Q_k$ relations at different global strains and we plot them for the 0.7~mm particles as an example in Fig.~\ref{fgr:j2_qk}(a). Here, we focus on the range of $Q_k$ where we have ample amount of data, and we normalize the bin-averaged $J_2$ by the average $J_2$ of all triangles at a specific $\epsilon/\epsilon_{SB}$, which is $\left<J_2\right>$. In the initial elastic regime at small $\epsilon/\epsilon_{SB}$, we found that the $J_2$--$Q_k$ relation is indeed different from the previously observed ``V-shape'' relation. Instead, $J_2$ is lowest at the smallest $Q_k$ and increases linearly with $Q_k$, especially for the few curves in the very beginning (light-colored). The "V-shape" relation is recovered in the period dominated by plastic rearrangements, especially for the last few curves. The transition between the two types of relation appears gradual, and this is essentially a dynamic-structure signature of the yield transition, which is yet another signal that coincides with strain localization. 

\begin{figure}[t]
\centering
  \includegraphics[width= 8.5 cm]{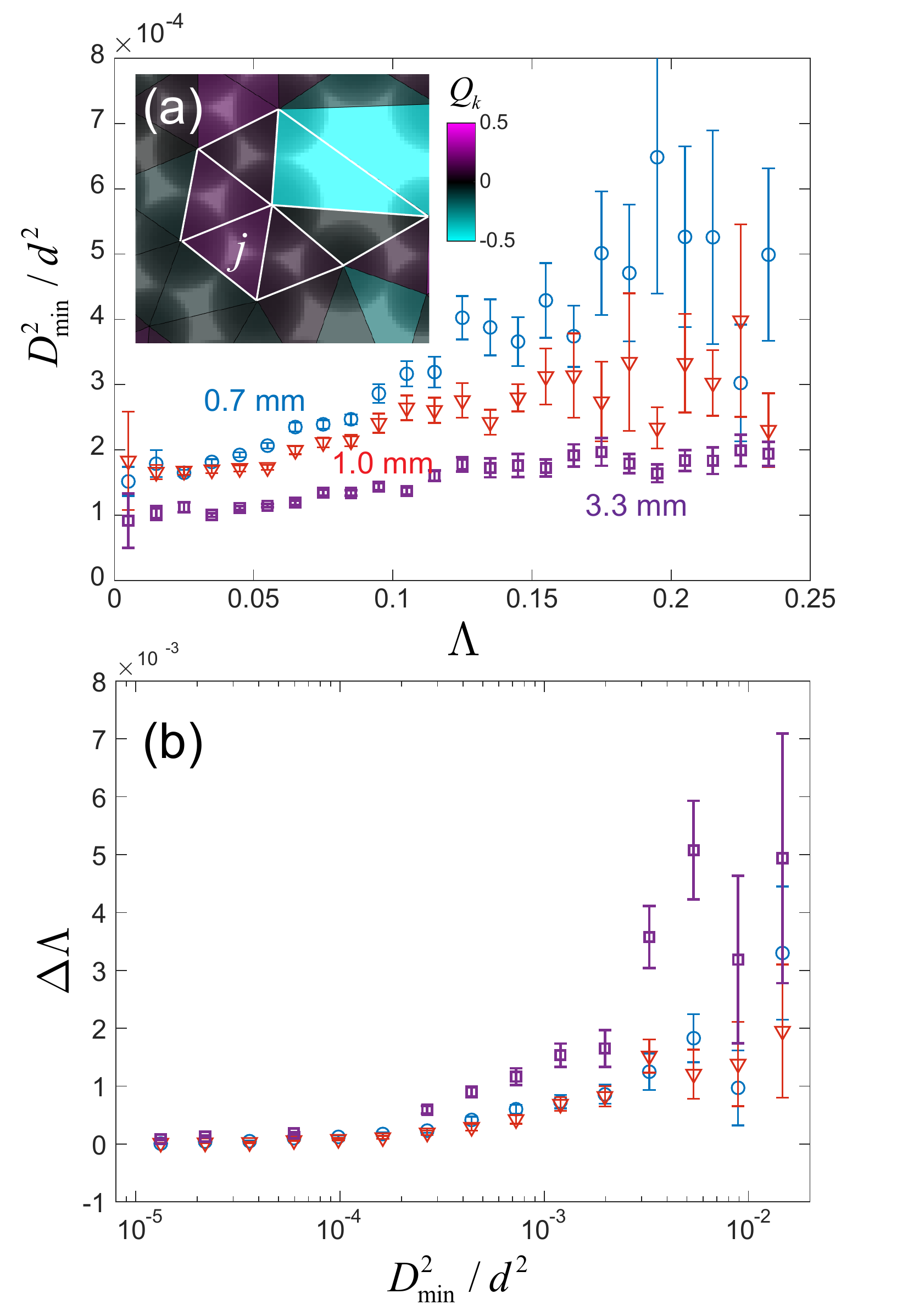}
  \caption{Interaction between local structure anisotropy and particle rearrangements. (a) Particle $D^2_{min}/d^2$ averaged by binning according to their $\Lambda$ values. (b) Local structure change $\Delta\Lambda$ averaged by binning according to their $D^2_{min}/d^2$ values. In both plots, error bars represent standard errors and symbols are blue circles for $d=0.7$~mm, orange triangles for $d=1.0$~mm, and purple squares for $d=3.3$~mm. }
  \label{fgr:qk_d2min}
\end{figure}

The difference in the $J_2$--$Q_k$ relation should come from the composition of the deformation itself. In the elastic period, $J_2$ is dominated by affine deformation, while in the later stage, $J_2$ is dominated by plastic rearrangements. To further investigate this difference, we compute the $J_2$--$Q_k$ relation in a second way. Instead of using triangles at a specific $\epsilon/\epsilon_{SB}$, we include all triangles collected within $0<\epsilon/\epsilon_{SB}<1$ and separate them into two groups: the ones that are near rearranging particles, and the ones far away from rearranging particles. Here we consider a particle rearranging when its $D^2_{min}/d^2>1\times10^{-4}$, a rather low threshold to make sure that we filter out all significant rearrangements and get truly affine deformation. This is also approximately a threshold where a rearrangement is starting to induce significant structural change (to be shown in Fig.~\ref{fgr:qk_d2min}(b)). We consider a triangle to be far away from a rearranging particle if it is not connected to either this particle or its first shell of neighbors (\textit{i.e.}, two layers away), and vice versa. 

The calculated $J_2$--$Q_k$ relations are shown in Fig.~\ref{fgr:j2_qk}(b) for all particle sizes, which show a clear difference between affine elastic deformation and plastic rearrangements. For triangles that are far away from rearrangements, the $J_2$--$Q_k$ relations for all three particle sizes show a linear increase of $J_2$ with $Q_k$, which is consistent with the early-stage results in Fig.~\ref{fgr:j2_qk}(a), confirming that it is indeed a dynamic-structure signature for elastic deformation. A possible explanation for $J_2$ being higher for more overpacked sites is that the capillary attractions for closer particles are stronger, so these particles can bear more elastic loading in the absence of rearrangements. On the other hand, plastic rearrangements appear to favor highly anisotropic sites, especially for highly underpacked sites with negative $Q_k$.\cite{harrington2020stagnant,harrington2018anisotropic} This is evident in the results of the $J_2$--$Q_k$ relations for the particles around rearrangements in Fig.~\ref{fgr:j2_qk}(b), which agree with the later-stage observations in Fig.~\ref{fgr:j2_qk}(a) that are dominated by plastic deformation. The rise of $J_2$ in the overpacked side ($Q_k>0$) of the ``V-shape'' appears to be more significant for smaller particles that are more ductile. For the 3.3~mm particles that have the shortest interaction range, it is possible that the strength of a very compact triangle is too strong for it to rearrange. In our previous studies of compression and penetration for particles of different shapes, dimers showed a substantially higher strength and poor ability to rearrange due to interlocking,\cite{harrington2020stagnant,harrington2018anisotropic} and they also do not have the upturn in the $J_2$--$Q_k$ relation. This similarity indicates that the upturn in the positive $Q_k$ side could serve as a signature for the degree of ease for certain particles to rearrange, and thus their ductility.

This preference for rearrangements to occur at highly structural anisotropic sites can be used to explain why a higher excess portion of such sites (Fig.~\ref{fgr:qk_phi}) leads to strain localization. To further explore the underlying mechanism, we directly quantify the interaction between structure and rearrangements, \textit{i.e.} $Q_k$ and $D^2_{min}$. To this end, we quantify the anisotropy of the local structure surrounding rearranging particles. The local anisotropy, $\Lambda$, is defined for each particle, where we collect all the triangles that shares a vertex with the particle, and compute the root mean square of their $Q_k$,
\begin{equation}
\label{eq:Lambda}
\Lambda = \left(\frac{1}{N}\sum_{j=1}^{N}Q_{k,j}^2\right)^{\frac{1}{2}}
\end{equation}
where $N$ is the total number of connected triangles and $Q_{k,j}$ is the $Q_k$ value for a connected triangle $j$.  This is demonstrated in Fig.~\ref{fgr:qk_d2min}(a) where the connected triangles are highlighted. By this construction, particles having a high packing anisotropy in their neighborhood, with either highly positive or negative $Q_k$, will have a high $\Lambda$, and vice versa. Note that the neighboring particles connected by these triangles are roughly also the same particles used for calculating $D^2_{min}$, which makes it appropriate to study the interaction between $\Lambda$ and $D^2_{min}$. Here, we consider particles that have $D^2_{min}/d^2>1\times10^{-5}$, and we calculate $\Lambda$ for a particle at the time instant when its $D^2_{min}$ reaches a temporal peak, $t_m$, noting that $D^2_{min}$ is largest when the entire rearrangement event is included in its calculation interval $\Delta t$. Thus $t_m$ is a reasonable starting time for a rearrangement. We then define the change of local structural anisotropy brought by a rearrangement as $\Delta\Lambda = \Lambda\left(t_m+\Delta t\right)-\Lambda\left(t_m\right)$. 

To further quantify the structural preference for rearrangements, we look at the average magnitude of a local rearrangement at a given structural anisotropy. The $D^2_{min}/d^2$ values of particles are binned according to their $\Lambda(t_m)$ values, and bin-averaged results for the three particle sizes are shown in Fig.~\ref{fgr:qk_d2min}(a). The results show that $D^2_{min}/d^2$ increases with $\Lambda$ for all three particle sizes, indicating that larger rearrangement events tend to occur at sites with higher local structural anisotropy. Also, under the same $\Lambda$, $D^2_{min}/d^2$ is larger for smaller particles that are more ductile. This result further indicates that smaller particles are more capable of rearranging to accommodate the global tensile deformation, which is an important factor of their ductility as discussed in Sec.~\ref{Sec3}. The structural change induced by a rearrangement event is studied by binning $\Delta\Lambda$ of particles according to their $D^2_{min}/d^2$; see Fig.~\ref{fgr:qk_d2min}(b). The results show that on average, for all particle sizes and all $D^2_{min}/d^2$ (except for extremely small $D^2_{min}/d^2$ that are below $1\times10^{-4}$),
$\Delta\Lambda$ is positive; therefore, the local structural anisotropy increases after rearrangements, and the magnitude of the increment is larger with larger $D^2_{min}/d^2$. The combination of Fig.~\ref{fgr:qk_d2min}(a) and~\ref{fgr:qk_d2min}(b) reveals why initial rearrangements are more likely to occur in regions with a higher excess portion $\Phi$, and how $\Delta\Phi$ is further enhanced by these rearrangements, resulting in strain localization. Thus this is a quantitative particle-level interpretation for the structure change observed in Fig.\ref{fgr:qk_phi}. Furthermore, figure~\ref{fgr:qk_d2min}(b) shows that under the same $D^2_{min}/d^2$, $\Delta\Lambda$ is larger for the 3.3~mm particles, indicating that the rearrangements in brittle materials can bring larger structural change and further facilitate strain localization. Under this mechanism, the initial larger $\Phi$ and $\Delta\Phi$ for more brittle materials (Figs.~\ref{fgr:qk_dist} and~\ref{fgr:qk_phi}), should also facilitate their strain localization process, reminiscent of studies showing dependence of ductility on preparation histories.\cite{lin2019distinguishing,ozawa2018random,gu2009ductile} 

\subsection{Structure-dynamic relations during failure}

Finally, we explore the structure-dynamic relation during the failure process, which greatly influences the ductility of the particles. For the failure process, while $Q_k$ is still a proper quantity for studying the structure-dynamic relation and identify growing voids (with highly negative $Q_k$), we choose to directly use the area of the triangles, $A_k$, to represent the local structure. The reason is that $A_k$ is a dimensional quantity that can be compared to important length scales such as $d$ and $l_c$. For the choice of a quantity representing the dynamics where void growth is important, we use the volumetric strain rate, $\dot{\epsilon}_v = \dot{\epsilon}_{11} + \dot{\epsilon}_{22}$, and also normalize it by $v_t/d$. The late stage of deformation, $\epsilon>\epsilon_{SB}$, is highly localized and all large rearrangements is confined in the failure region (Fig.~\ref{fgr:d2map}). Thus, we focus on triangles in this region, which can be found by selecting triangles around particles with high $D^2_{min}$. We set the threshold to be $D^2_{min}/d^2>1\times10^{-3}$ and the search radius to be two layers around these particles so that roughly all triangles in this zone are included. 

\begin{figure}[t]
\centering
  \includegraphics[width= 8.9 cm]{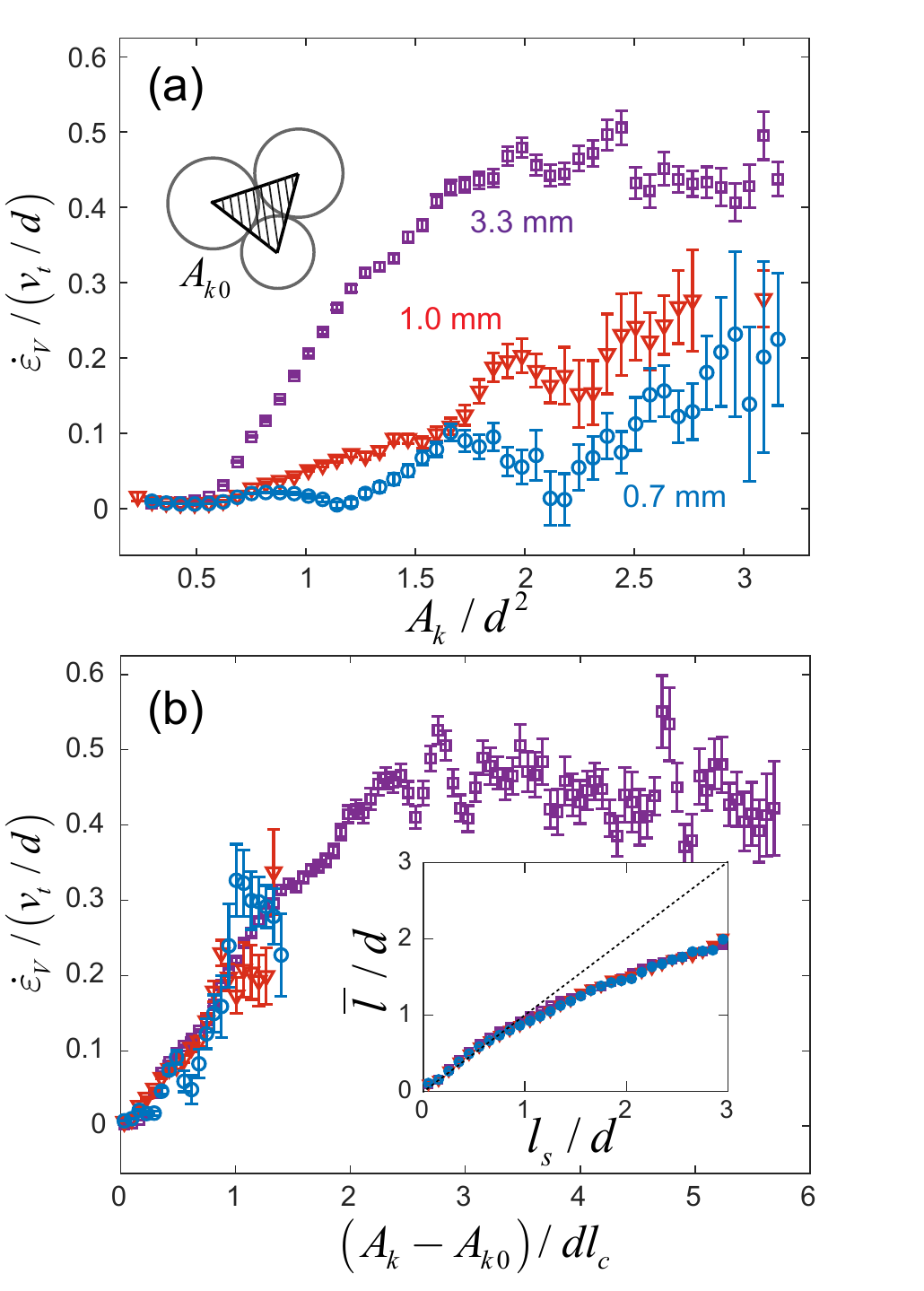}
  \caption{Structure-dynamic relation during failure. (a) Normalized triangle volumetric strain rate, $\dot{\epsilon}_v/(v_t/d)$, averaged by binning according to their normalized areas, $A_k/d^2$. (b) $\dot{\epsilon}_v/(v_t/d)$ binned according to $(A_k-A_{k0})/dl_c$. Inset: the average pairwise separation distance $\bar{l}/d$ vs. $l_s/d$ for the three particle sizes. In all plots, error bars represent standard errors and symbols are blue circles for $d=0.7$~mm, orange triangles for $d=1.0$~mm, and purple squares for $d=3.3$~mm.}
  \label{fgr:area_ev}
\end{figure}

The structure-dynamics relation between the triangle areas and the volumetric strain rate is shown in Fig.~\ref{fgr:area_ev}. For each particle size, we bin $\dot{\epsilon}_v/\left(v_t/d\right)$ by $A_k/d^2$ and show the bin-averaged values in Fig.~\ref{fgr:area_ev}(a). The results show that the averaged $\dot{\epsilon}_v/\left(v_t/d\right)$ is positive, and triangles with larger areas tend to have larger volumetric strain rates. This trend works in favor of the growth of larger voids during failure, and the differential growth rate of $A_k$ can further increase the local packing anisotropy, which is consistent with the $Q_k$-based measure in Fig.~\ref{fgr:qk_phi}. Moreover, the increase rate of $\dot{\epsilon}_v/\left(v_t/d\right)$ with $A_k/d^2$ is significantly larger for larger particles, showing that the void growth for brittle failure is more dramatic. For ductile failure, while $\dot{\epsilon}_v/\left(v_t/d\right)$ for the 0.7~mm particles is relatively low, the deviatoric strain rate $J_2$ can actually be higher than $J_2$ for the brittle materials, see Fig.~\ref{fgr:j2map}. This observation agrees with Fig.~\ref{fgr:qk_d2min} that $\Delta\Lambda$ is lower for more ductile particles for the same $D^2_{min}/d^2$. These results quantitatively show that ductile materials can have larger-scale particle rearrangements while avoiding any significant void growth during failure. 

The initial plateau at small $A_k/d^2$ in Fig.~\ref{fgr:area_ev}(a) corresponds to the regime where particles are in close contact. The plateau exists because $A_k/d^2$ at close contact is not a unique value due to the polydispersity of the particle diameters. To correct this, we subtract the triangle area corresponding to close contact, $A_{k0}$, which is calculated using the actual diameter of the constituting particles, as sketched in Fig.~\ref{fgr:area_ev}. In this way, $A_k-A_{k0}$ is the actual area that corresponds to the separation of particles for a triangle. Upon further examination, we found that the increase rate of $\dot{\epsilon}_v/\left(v_t/d\right)$ with $(A_k-A_{k0})/d^2$ is proportional to $d$, which motivated us to scale $(A_k-A_{k0})/d^2$ by $l_c/d$. Figure~\ref{fgr:area_ev}(b) shows the results of $\dot{\epsilon}_v/\left(v_t/d\right)$ that is bin-averaged according to $(A_k-A_{k0})/dl_c$. In this way, data for the three particle sizes collapses, showing a linear initial increase followed by a plateau at large areas. 

One way to rationalize this new scaling for the opened area is that during uniaxial tensile deformation, the triangles mainly expand in one direction, while the separation in the other direction should remain at a characteristic length of $d$. Thus $(A_k-A_{k0})/d$ gives a length scale for the separation of the particles in the expanding direction, $l_s$, which can be compared to $l_c$. Within $l_s/l_c\approx2$, the increase of $l_s$ results in faster expansion of the triangles. The upper limit of this regime, $l_s/l_c\approx2$, somehow agrees with the decay range of the capillary attraction $f_c$ as shown in Fig.~\ref{fgr:capillary}(b). The inset in Fig.~\ref{fgr:area_ev}(b) shows how this length scale compares with the normalized average pairwise separation distance of the three particles in a triangle, $\bar{l}/d$. The results show that the area-based length scale, $l_s$, is comparable to $\bar{l}$ in magnitude, which in turn indicates that $l_s$ is comparable to $l$ in Fig.~\ref{fgr:capillary}(b). For $l_s/l_c>2$, a second regime is reached where $\dot{\epsilon}_v$ is roughly proportional to $v_t/d$, and $l_s/l_c$ no longer has an influence over it, suggesting that the particles have broken free from the capillary attraction in the expansion direction. This plateau in the volumetric strain rate also means that the area increase rate, $dA_k/dt$, is proportional to $A_k$, thus these huge triangles should be growing exponentially in time, a rather drastic growth mode that facilitates failure.  Only the 3.3~mm particles reached this regime. Again, brittle and ductile failure are distinguished via the structure-dynamic relation, the scaling based on $l_s$ reveals how ductility is tuned by the interaction range $l_c/d$ at the particle level.

\section{Conclusions}
\label{Sec5}

We have shown that a model disordered solid made of a particle raft can experience elastic deformation, strain localization, and failure during quasi-static tensile tests. The ductility of the material can be tuned by using different particle sizes, which in turn controls the interaction range for the capillary attractions. Smaller particles with longer interaction ranges are more ductile and can endure larger global tensile deformation without forming larger voids and fractures. Larger particles with shorter interaction ranges fail in a brittle way with a fracture forming after relatively small global tensile deformation. Distinct local structure-dynamic relations were found between elastic deformation and plastic rearrangements, with the latter being responsible for inducing structural changes and strain localization. The excess portion of local sites with high structural anisotropy was found to be higher for strain localized regions and also for more brittle materials. These sites are more prone to having local particle rearrangements, which can in turn further raise the structural anisotropy, forming a mechanism that leads to strain localization. During strain localization and failure, smaller particles can organize into larger rearrangements while keeping voids from growing, resulting in ductile behaviors. On the contrary, rearrangement in larger particles relies more on the highly underpacked sites, and a larger differential void growth was found for these particles, leading to brittle behaviors.

The experimental method developed in this work is useful for studying the influences of particle-level features on local structures and dynamics as well as emergent system-scale behaviors such as the shear band formation, despite the relatively small system size.
While this study demonstrates how the particle interaction range controls ductility as previous simulations predicted,\cite{falk1999molecular,dauchot2011athermal,babu2016excess,lin2019distinguishing} more mechanisms can also be tested using this apparatus. It is certainly possible to tune the capillary interaction more finely by adjusting particle/fluid density ratio, contact angle, surface tension, or even replace the liquid surface tension with a long-range elastic tension by covering particles with a thin elastic film.\cite{lee2020cheerios} We can also test the influence of surface friction,\cite{karimi2019plastic} deformability,\cite{boromand2019role} and particle shape\cite{zhang2013using} by using bubbles or 3D-printed particles. Including thermal noise by mechanically vibrating the liquid surface to study the effect of temperature and quenching\cite{ozawa2018random,lin2019distinguishing} is also possible. These methods can be used for exploring, designing, and optimizing ductility and other mechanical properties of disordered materials.

While revealing important features in the brittle-to-ductile transition, the experimental results here are also informative for understanding and modelling plastic deformation of disordered solids. Firstly, the distinct structure-dynamic relations observed between elastic and plastic deformation support the general elasto-plastic models which treat the deformation of disordered materials as plastic rearranging sites connected by long-range elastic strains.\cite{nicolas2018deformation} The non-trivial slope in the $J_2$--$Q_k$ relation for elastic deformation suggests that the structural anisotropy could play an important role in elastic interactions. Secondly, the $Q_k$ based structure function, $\Lambda$, developed in this study has good correlations with plastic rearrangements and has a clear physical meaning. Thus it can be used to compare with other structure functions that accurately predict structural defects but have more elusive physical meanings.\cite{cubuk2017structure,cubuk2015identifying} Lastly, the rapid growth of voids plays a critical role in brittle failure, and could be studied to test if the void growth rate transition in Fig.~\ref{fgr:area_ev}(b) can be reached for ductile materials. It is possible that certain ductile-to-brittle transitions, due to factors such as the system size,\cite{guo2007tensile} is related to whether the plateau corresponding to an exponential void growth is reached.



\section*{Conflicts of interest}
There are no conflicts to declare.

\section*{Acknowledgements}
The authors would like to acknowledge the funding from the National Science Foundation grant MRSEC/DMR-1720530. We also thank Andrea~J.~Liu, Ge Zhang, and Robert~A.~Riggleman for helpful suggestions, as well as Douglas~J.~Jerolmack's group for the help with measuring particle sizes. 


\balance


\bibliography{rsc} 

\providecommand*{\mcitethebibliography}{\thebibliography}
\csname @ifundefined\endcsname{endmcitethebibliography}
{\let\endmcitethebibliography\endthebibliography}{}
\begin{mcitethebibliography}{73}
\providecommand*{\natexlab}[1]{#1}
\providecommand*{\mciteSetBstSublistMode}[1]{}
\providecommand*{\mciteSetBstMaxWidthForm}[2]{}
\providecommand*{\mciteBstWouldAddEndPuncttrue}
  {\def\EndOfBibitem{\unskip.}}
\providecommand*{\mciteBstWouldAddEndPunctfalse}
  {\let\EndOfBibitem\relax}
\providecommand*{\mciteSetBstMidEndSepPunct}[3]{}
\providecommand*{\mciteSetBstSublistLabelBeginEnd}[3]{}
\providecommand*{\EndOfBibitem}{}
\mciteSetBstSublistMode{f}
\mciteSetBstMaxWidthForm{subitem}
{(\emph{\alph{mcitesubitemcount}})}
\mciteSetBstSublistLabelBeginEnd{\mcitemaxwidthsubitemform\space}
{\relax}{\relax}

\bibitem[Chen(2008)]{chen2008mechanical}
M.~Chen, \emph{Annu. Rev. Mater. Res.}, 2008, \textbf{38}, 445--469\relax
\mciteBstWouldAddEndPuncttrue
\mciteSetBstMidEndSepPunct{\mcitedefaultmidpunct}
{\mcitedefaultendpunct}{\mcitedefaultseppunct}\relax
\EndOfBibitem
\bibitem[Greer \emph{et~al.}(2013)Greer, Cheng, and Ma]{greer2013shear}
A.~Greer, Y.~Cheng and E.~Ma, \emph{Mater. Sci. Eng. R Rep.}, 2013,
  \textbf{74}, 71--132\relax
\mciteBstWouldAddEndPuncttrue
\mciteSetBstMidEndSepPunct{\mcitedefaultmidpunct}
{\mcitedefaultendpunct}{\mcitedefaultseppunct}\relax
\EndOfBibitem
\bibitem[Lan \emph{et~al.}(2019)Lan, Chen, and Macciotta]{lan2019universal}
H.~Lan, J.~Chen and R.~Macciotta, \emph{Sci. Rep.}, 2019, \textbf{9},
  1--9\relax
\mciteBstWouldAddEndPuncttrue
\mciteSetBstMidEndSepPunct{\mcitedefaultmidpunct}
{\mcitedefaultendpunct}{\mcitedefaultseppunct}\relax
\EndOfBibitem
\bibitem[Zhang \emph{et~al.}(2013)Zhang, Feng, Zeravcic, Brugarolas, Liu, and
  Lee]{zhang2013using}
L.~Zhang, G.~Feng, Z.~Zeravcic, T.~Brugarolas, A.~J. Liu and D.~Lee, \emph{ACS
  nano}, 2013, \textbf{7}, 8043--8050\relax
\mciteBstWouldAddEndPuncttrue
\mciteSetBstMidEndSepPunct{\mcitedefaultmidpunct}
{\mcitedefaultendpunct}{\mcitedefaultseppunct}\relax
\EndOfBibitem
\bibitem[Manning \emph{et~al.}(2007)Manning, Langer, and
  Carlson]{manning2007strain}
M.~L. Manning, J.~S. Langer and J.~Carlson, \emph{Phys. Rev. E}, 2007,
  \textbf{76}, 056106\relax
\mciteBstWouldAddEndPuncttrue
\mciteSetBstMidEndSepPunct{\mcitedefaultmidpunct}
{\mcitedefaultendpunct}{\mcitedefaultseppunct}\relax
\EndOfBibitem
\bibitem[Le~Bouil \emph{et~al.}(2014)Le~Bouil, Amon, McNamara, and
  Crassous]{le2014emergence}
A.~Le~Bouil, A.~Amon, S.~McNamara and J.~Crassous, \emph{Phys. Rev. Lett.},
  2014, \textbf{112}, 246001\relax
\mciteBstWouldAddEndPuncttrue
\mciteSetBstMidEndSepPunct{\mcitedefaultmidpunct}
{\mcitedefaultendpunct}{\mcitedefaultseppunct}\relax
\EndOfBibitem
\bibitem[Ma and Elbanna(2018)]{ma2018strain}
X.~Ma and A.~Elbanna, \emph{Phys. Rev. E}, 2018, \textbf{98}, 022906\relax
\mciteBstWouldAddEndPuncttrue
\mciteSetBstMidEndSepPunct{\mcitedefaultmidpunct}
{\mcitedefaultendpunct}{\mcitedefaultseppunct}\relax
\EndOfBibitem
\bibitem[Lin and Riggleman(2019)]{lin2019distinguishing}
E.~Y. Lin and R.~A. Riggleman, \emph{Soft Matter}, 2019, \textbf{15},
  6589--6595\relax
\mciteBstWouldAddEndPuncttrue
\mciteSetBstMidEndSepPunct{\mcitedefaultmidpunct}
{\mcitedefaultendpunct}{\mcitedefaultseppunct}\relax
\EndOfBibitem
\bibitem[Ivancic and Riggleman(2019)]{ivancic2019identifying}
R.~J. Ivancic and R.~A. Riggleman, \emph{Soft Matter}, 2019, \textbf{15},
  4548--4561\relax
\mciteBstWouldAddEndPuncttrue
\mciteSetBstMidEndSepPunct{\mcitedefaultmidpunct}
{\mcitedefaultendpunct}{\mcitedefaultseppunct}\relax
\EndOfBibitem
\bibitem[Wang \emph{et~al.}(2006)Wang, Krishan, and Dennin]{wang2006impact}
Y.~Wang, K.~Krishan and M.~Dennin, \emph{Phys. Rev. E}, 2006, \textbf{73},
  031401\relax
\mciteBstWouldAddEndPuncttrue
\mciteSetBstMidEndSepPunct{\mcitedefaultmidpunct}
{\mcitedefaultendpunct}{\mcitedefaultseppunct}\relax
\EndOfBibitem
\bibitem[Katgert \emph{et~al.}(2008)Katgert, M{\"o}bius, and van
  Hecke]{katgert2008rate}
G.~Katgert, M.~E. M{\"o}bius and M.~van Hecke, \emph{Phys. Rev. Lett.}, 2008,
  \textbf{101}, 058301\relax
\mciteBstWouldAddEndPuncttrue
\mciteSetBstMidEndSepPunct{\mcitedefaultmidpunct}
{\mcitedefaultendpunct}{\mcitedefaultseppunct}\relax
\EndOfBibitem
\bibitem[Tapia \emph{et~al.}(2013)Tapia, Esp{\'\i}ndola, Hamm, and
  Melo]{tapia2013effect}
F.~Tapia, D.~Esp{\'\i}ndola, E.~Hamm and F.~Melo, \emph{Phys. Rev. E}, 2013,
  \textbf{87}, 014201\relax
\mciteBstWouldAddEndPuncttrue
\mciteSetBstMidEndSepPunct{\mcitedefaultmidpunct}
{\mcitedefaultendpunct}{\mcitedefaultseppunct}\relax
\EndOfBibitem
\bibitem[Schall and van Hecke(2010)]{schall2010shear}
P.~Schall and M.~van Hecke, \emph{Annu. Rev. Fluid Mech.}, 2010, \textbf{42},
  67--88\relax
\mciteBstWouldAddEndPuncttrue
\mciteSetBstMidEndSepPunct{\mcitedefaultmidpunct}
{\mcitedefaultendpunct}{\mcitedefaultseppunct}\relax
\EndOfBibitem
\bibitem[Kawamoto \emph{et~al.}(2018)Kawamoto, And{\`o}, Viggiani, and
  Andrade]{kawamoto2018all}
R.~Kawamoto, E.~And{\`o}, G.~Viggiani and J.~E. Andrade, \emph{J. Mech. Phys.
  Solids}, 2018, \textbf{111}, 375--392\relax
\mciteBstWouldAddEndPuncttrue
\mciteSetBstMidEndSepPunct{\mcitedefaultmidpunct}
{\mcitedefaultendpunct}{\mcitedefaultseppunct}\relax
\EndOfBibitem
\bibitem[Falk and Langer(1998)]{falk1998dynamics}
M.~L. Falk and J.~S. Langer, \emph{Phys. Rev. E}, 1998, \textbf{57}, 7192\relax
\mciteBstWouldAddEndPuncttrue
\mciteSetBstMidEndSepPunct{\mcitedefaultmidpunct}
{\mcitedefaultendpunct}{\mcitedefaultseppunct}\relax
\EndOfBibitem
\bibitem[Nicolas \emph{et~al.}(2018)Nicolas, Ferrero, Martens, and
  Barrat]{nicolas2018deformation}
A.~Nicolas, E.~E. Ferrero, K.~Martens and J.-L. Barrat, \emph{Rev. Mod. Phys.},
  2018, \textbf{90}, 045006\relax
\mciteBstWouldAddEndPuncttrue
\mciteSetBstMidEndSepPunct{\mcitedefaultmidpunct}
{\mcitedefaultendpunct}{\mcitedefaultseppunct}\relax
\EndOfBibitem
\bibitem[Cubuk \emph{et~al.}(2017)Cubuk, Ivancic, Schoenholz, Strickland, Basu,
  Davidson, Fontaine, Hor, Huang, Jiang,\emph{et~al.}]{cubuk2017structure}
E.~D. Cubuk, R.~Ivancic, S.~S. Schoenholz, D.~Strickland, A.~Basu, Z.~Davidson,
  J.~Fontaine, J.~L. Hor, Y.-R. Huang, Y.~Jiang \emph{et~al.}, \emph{Science},
  2017, \textbf{358}, 1033--1037\relax
\mciteBstWouldAddEndPuncttrue
\mciteSetBstMidEndSepPunct{\mcitedefaultmidpunct}
{\mcitedefaultendpunct}{\mcitedefaultseppunct}\relax
\EndOfBibitem
\bibitem[Dansereau \emph{et~al.}(2019)Dansereau, D{\'e}mery, Berthier, Weiss,
  and Ponson]{dansereau2019collective}
V.~Dansereau, V.~D{\'e}mery, E.~Berthier, J.~Weiss and L.~Ponson, \emph{Phys.
  Rev. Lett.}, 2019, \textbf{122}, 085501\relax
\mciteBstWouldAddEndPuncttrue
\mciteSetBstMidEndSepPunct{\mcitedefaultmidpunct}
{\mcitedefaultendpunct}{\mcitedefaultseppunct}\relax
\EndOfBibitem
\bibitem[Karimi and Barrat(2018)]{karimi2018correlation}
K.~Karimi and J.-L. Barrat, \emph{Sci. Rep.}, 2018, \textbf{8}, 1--10\relax
\mciteBstWouldAddEndPuncttrue
\mciteSetBstMidEndSepPunct{\mcitedefaultmidpunct}
{\mcitedefaultendpunct}{\mcitedefaultseppunct}\relax
\EndOfBibitem
\bibitem[Dasgupta \emph{et~al.}(2012)Dasgupta, Hentschel, and
  Procaccia]{dasgupta2012microscopic}
R.~Dasgupta, H.~G.~E. Hentschel and I.~Procaccia, \emph{Phys. Rev. Lett.},
  2012, \textbf{109}, 255502\relax
\mciteBstWouldAddEndPuncttrue
\mciteSetBstMidEndSepPunct{\mcitedefaultmidpunct}
{\mcitedefaultendpunct}{\mcitedefaultseppunct}\relax
\EndOfBibitem
\bibitem[Falk and Langer(2011)]{falk2011deformation}
M.~L. Falk and J.~S. Langer, \emph{Annu. Rev. Condens. Matter Phys.}, 2011,
  \textbf{2}, 353--373\relax
\mciteBstWouldAddEndPuncttrue
\mciteSetBstMidEndSepPunct{\mcitedefaultmidpunct}
{\mcitedefaultendpunct}{\mcitedefaultseppunct}\relax
\EndOfBibitem
\bibitem[Sollich \emph{et~al.}(1997)Sollich, Lequeux, H{\'e}braud, and
  Cates]{sollich1997rheology}
P.~Sollich, F.~Lequeux, P.~H{\'e}braud and M.~E. Cates, \emph{Phys. Rev.
  Lett.}, 1997, \textbf{78}, 2020\relax
\mciteBstWouldAddEndPuncttrue
\mciteSetBstMidEndSepPunct{\mcitedefaultmidpunct}
{\mcitedefaultendpunct}{\mcitedefaultseppunct}\relax
\EndOfBibitem
\bibitem[Lauridsen \emph{et~al.}(2002)Lauridsen, Twardos, and
  Dennin]{lauridsen2002shear}
J.~Lauridsen, M.~Twardos and M.~Dennin, \emph{Phys. Rev. Lett.}, 2002,
  \textbf{89}, 098303\relax
\mciteBstWouldAddEndPuncttrue
\mciteSetBstMidEndSepPunct{\mcitedefaultmidpunct}
{\mcitedefaultendpunct}{\mcitedefaultseppunct}\relax
\EndOfBibitem
\bibitem[Harrington and Durian(2018)]{harrington2018anisotropic}
M.~Harrington and D.~J. Durian, \emph{Phys. Rev. E}, 2018, \textbf{97},
  012904\relax
\mciteBstWouldAddEndPuncttrue
\mciteSetBstMidEndSepPunct{\mcitedefaultmidpunct}
{\mcitedefaultendpunct}{\mcitedefaultseppunct}\relax
\EndOfBibitem
\bibitem[Kuo and Dennin(2012)]{kuo2012scaling}
C.-C. Kuo and M.~Dennin, \emph{J. Rheol.}, 2012, \textbf{56}, 527--541\relax
\mciteBstWouldAddEndPuncttrue
\mciteSetBstMidEndSepPunct{\mcitedefaultmidpunct}
{\mcitedefaultendpunct}{\mcitedefaultseppunct}\relax
\EndOfBibitem
\bibitem[Karimi \emph{et~al.}(2019)Karimi, Amitrano, and
  Weiss]{karimi2019plastic}
K.~Karimi, D.~Amitrano and J.~Weiss, \emph{Phys. Rev. E}, 2019, \textbf{100},
  012908\relax
\mciteBstWouldAddEndPuncttrue
\mciteSetBstMidEndSepPunct{\mcitedefaultmidpunct}
{\mcitedefaultendpunct}{\mcitedefaultseppunct}\relax
\EndOfBibitem
\bibitem[Gu \emph{et~al.}(2009)Gu, Poon, Shiflet, and
  Lewandowski]{gu2009ductile}
X.~Gu, S.~J. Poon, G.~J. Shiflet and J.~Lewandowski, \emph{Scr. Mater.}, 2009,
  \textbf{60}, 1027--1030\relax
\mciteBstWouldAddEndPuncttrue
\mciteSetBstMidEndSepPunct{\mcitedefaultmidpunct}
{\mcitedefaultendpunct}{\mcitedefaultseppunct}\relax
\EndOfBibitem
\bibitem[Ozawa \emph{et~al.}(2018)Ozawa, Berthier, Biroli, Rosso, and
  Tarjus]{ozawa2018random}
M.~Ozawa, L.~Berthier, G.~Biroli, A.~Rosso and G.~Tarjus, \emph{Proc. Natl.
  Acad. Sci.}, 2018, \textbf{115}, 6656--6661\relax
\mciteBstWouldAddEndPuncttrue
\mciteSetBstMidEndSepPunct{\mcitedefaultmidpunct}
{\mcitedefaultendpunct}{\mcitedefaultseppunct}\relax
\EndOfBibitem
\bibitem[Guo \emph{et~al.}(2007)Guo, Yan, Wang, Tan, Zhang, Sui, and
  Ma]{guo2007tensile}
H.~Guo, P.~Yan, Y.~Wang, J.~Tan, Z.~Zhang, M.~Sui and E.~Ma, \emph{Nature
  Mat.}, 2007, \textbf{6}, 735--739\relax
\mciteBstWouldAddEndPuncttrue
\mciteSetBstMidEndSepPunct{\mcitedefaultmidpunct}
{\mcitedefaultendpunct}{\mcitedefaultseppunct}\relax
\EndOfBibitem
\bibitem[Sopu \emph{et~al.}(2016)Sopu, Foroughi, Stoica, and
  Eckert]{sopu2016brittle}
D.~Sopu, A.~Foroughi, M.~Stoica and J.~Eckert, \emph{Nano Lett.}, 2016,
  \textbf{16}, 4467--4471\relax
\mciteBstWouldAddEndPuncttrue
\mciteSetBstMidEndSepPunct{\mcitedefaultmidpunct}
{\mcitedefaultendpunct}{\mcitedefaultseppunct}\relax
\EndOfBibitem
\bibitem[Cho \emph{et~al.}(2019)Cho, Lu, Howard, Adams, and
  Datta]{cho2019crack}
H.~J. Cho, N.~B. Lu, M.~P. Howard, R.~A. Adams and S.~S. Datta, \emph{Soft
  Matter}, 2019, \textbf{15}, 4689--4702\relax
\mciteBstWouldAddEndPuncttrue
\mciteSetBstMidEndSepPunct{\mcitedefaultmidpunct}
{\mcitedefaultendpunct}{\mcitedefaultseppunct}\relax
\EndOfBibitem
\bibitem[Shi(2010)]{shi2010size}
Y.~Shi, \emph{Appl. Phys. Lett.}, 2010, \textbf{96}, 121909\relax
\mciteBstWouldAddEndPuncttrue
\mciteSetBstMidEndSepPunct{\mcitedefaultmidpunct}
{\mcitedefaultendpunct}{\mcitedefaultseppunct}\relax
\EndOfBibitem
\bibitem[Falk(1999)]{falk1999molecular}
M.~Falk, \emph{Physical Review B}, 1999, \textbf{60}, 7062\relax
\mciteBstWouldAddEndPuncttrue
\mciteSetBstMidEndSepPunct{\mcitedefaultmidpunct}
{\mcitedefaultendpunct}{\mcitedefaultseppunct}\relax
\EndOfBibitem
\bibitem[Dauchot \emph{et~al.}(2011)Dauchot, Karmakar, Procaccia, and
  Zylberg]{dauchot2011athermal}
O.~Dauchot, S.~Karmakar, I.~Procaccia and J.~Zylberg, \emph{Phys. Rev. E},
  2011, \textbf{84}, 046105\relax
\mciteBstWouldAddEndPuncttrue
\mciteSetBstMidEndSepPunct{\mcitedefaultmidpunct}
{\mcitedefaultendpunct}{\mcitedefaultseppunct}\relax
\EndOfBibitem
\bibitem[Babu \emph{et~al.}(2016)Babu, Mondal, Sengupta, and
  Karmakar]{babu2016excess}
J.~S. Babu, C.~Mondal, S.~Sengupta and S.~Karmakar, \emph{Soft Matter}, 2016,
  \textbf{12}, 1210--1218\relax
\mciteBstWouldAddEndPuncttrue
\mciteSetBstMidEndSepPunct{\mcitedefaultmidpunct}
{\mcitedefaultendpunct}{\mcitedefaultseppunct}\relax
\EndOfBibitem
\bibitem[Schmeink \emph{et~al.}(2017)Schmeink, Goehring, and
  Hemmerle]{schmeink2017fracture}
A.~Schmeink, L.~Goehring and A.~Hemmerle, \emph{Soft Matter}, 2017,
  \textbf{13}, 1040--1047\relax
\mciteBstWouldAddEndPuncttrue
\mciteSetBstMidEndSepPunct{\mcitedefaultmidpunct}
{\mcitedefaultendpunct}{\mcitedefaultseppunct}\relax
\EndOfBibitem
\bibitem[Hemmerle \emph{et~al.}(2016)Hemmerle, Schr{\"o}ter, and
  Goehring]{hemmerle2016cohesive}
A.~Hemmerle, M.~Schr{\"o}ter and L.~Goehring, \emph{Sci. Rep.}, 2016,
  \textbf{6}, 35650\relax
\mciteBstWouldAddEndPuncttrue
\mciteSetBstMidEndSepPunct{\mcitedefaultmidpunct}
{\mcitedefaultendpunct}{\mcitedefaultseppunct}\relax
\EndOfBibitem
\bibitem[Hor \emph{et~al.}(2017)Hor, Jiang, Ring, Riggleman, Turner, and
  Lee]{hor2017nanoporous}
J.~L. Hor, Y.~Jiang, D.~J. Ring, R.~A. Riggleman, K.~T. Turner and D.~Lee,
  \emph{ACS nano}, 2017, \textbf{11}, 3229--3236\relax
\mciteBstWouldAddEndPuncttrue
\mciteSetBstMidEndSepPunct{\mcitedefaultmidpunct}
{\mcitedefaultendpunct}{\mcitedefaultseppunct}\relax
\EndOfBibitem
\bibitem[Jiang \emph{et~al.}(2018)Jiang, Hor, Lee, and
  Turner]{jiang2018toughening}
Y.~Jiang, J.~L. Hor, D.~Lee and K.~T. Turner, \emph{ACS Appl. Mater.
  Interfaces}, 2018, \textbf{10}, 44011--44017\relax
\mciteBstWouldAddEndPuncttrue
\mciteSetBstMidEndSepPunct{\mcitedefaultmidpunct}
{\mcitedefaultendpunct}{\mcitedefaultseppunct}\relax
\EndOfBibitem
\bibitem[Nicolson(1949)]{nicolson1949interaction}
M.~Nicolson, Math. Proc. Camb. Philos. Soc., 1949, pp. 288--295\relax
\mciteBstWouldAddEndPuncttrue
\mciteSetBstMidEndSepPunct{\mcitedefaultmidpunct}
{\mcitedefaultendpunct}{\mcitedefaultseppunct}\relax
\EndOfBibitem
\bibitem[Kralchevsky and Nagayama(2000)]{kralchevsky2000capillary}
P.~A. Kralchevsky and K.~Nagayama, \emph{Adv. Colloid Interface Sci.}, 2000,
  \textbf{85}, 145--192\relax
\mciteBstWouldAddEndPuncttrue
\mciteSetBstMidEndSepPunct{\mcitedefaultmidpunct}
{\mcitedefaultendpunct}{\mcitedefaultseppunct}\relax
\EndOfBibitem
\bibitem[Singh and Joseph(2005)]{singh2005fluid}
P.~Singh and D.~Joseph, \emph{J. Fluid Mech.}, 2005, \textbf{530}, 31--80\relax
\mciteBstWouldAddEndPuncttrue
\mciteSetBstMidEndSepPunct{\mcitedefaultmidpunct}
{\mcitedefaultendpunct}{\mcitedefaultseppunct}\relax
\EndOfBibitem
\bibitem[Dalbe \emph{et~al.}(2011)Dalbe, Cosic, Berhanu, and
  Kudrolli]{dalbe2011aggregation}
M.-J. Dalbe, D.~Cosic, M.~Berhanu and A.~Kudrolli, \emph{Phys. Rev. E}, 2011,
  \textbf{83}, 051403\relax
\mciteBstWouldAddEndPuncttrue
\mciteSetBstMidEndSepPunct{\mcitedefaultmidpunct}
{\mcitedefaultendpunct}{\mcitedefaultseppunct}\relax
\EndOfBibitem
\bibitem[Ho \emph{et~al.}(2019)Ho, Pucci, and Harris]{ho2019direct}
I.~Ho, G.~Pucci and D.~M. Harris, \emph{Phys. Rev. Lett.}, 2019, \textbf{123},
  254502\relax
\mciteBstWouldAddEndPuncttrue
\mciteSetBstMidEndSepPunct{\mcitedefaultmidpunct}
{\mcitedefaultendpunct}{\mcitedefaultseppunct}\relax
\EndOfBibitem
\bibitem[Bragg and Nye(1947)]{bragg1947dynamical}
W.~L. Bragg and J.~Nye, \emph{Proc. R. Soc. A}, 1947, \textbf{190},
  474--481\relax
\mciteBstWouldAddEndPuncttrue
\mciteSetBstMidEndSepPunct{\mcitedefaultmidpunct}
{\mcitedefaultendpunct}{\mcitedefaultseppunct}\relax
\EndOfBibitem
\bibitem[Mazuyer \emph{et~al.}(1989)Mazuyer, Georges, and
  Cambou]{mazuyer1989shear}
D.~Mazuyer, J.~Georges and B.~Cambou, \emph{J. Phys.}, 1989, \textbf{50},
  1057--1067\relax
\mciteBstWouldAddEndPuncttrue
\mciteSetBstMidEndSepPunct{\mcitedefaultmidpunct}
{\mcitedefaultendpunct}{\mcitedefaultseppunct}\relax
\EndOfBibitem
\bibitem[Kozlowski \emph{et~al.}(2019)Kozlowski, Carlevaro, Daniels, Kondic,
  Pugnaloni, Socolar, Zheng, and Behringer]{kozlowski2019dynamics}
R.~Kozlowski, C.~M. Carlevaro, K.~E. Daniels, L.~Kondic, L.~A. Pugnaloni, J.~E.
  Socolar, H.~Zheng and R.~P. Behringer, \emph{Phys. Rev. E}, 2019,
  \textbf{100}, 032905\relax
\mciteBstWouldAddEndPuncttrue
\mciteSetBstMidEndSepPunct{\mcitedefaultmidpunct}
{\mcitedefaultendpunct}{\mcitedefaultseppunct}\relax
\EndOfBibitem
\bibitem[Liu \emph{et~al.}(2018)Liu, Sharifi-Mood, and Stebe]{liu2018capillary}
I.~B. Liu, N.~Sharifi-Mood and K.~J. Stebe, \emph{Annu. Rev. Condens. Matter
  Phys.}, 2018, \textbf{9}, 283--305\relax
\mciteBstWouldAddEndPuncttrue
\mciteSetBstMidEndSepPunct{\mcitedefaultmidpunct}
{\mcitedefaultendpunct}{\mcitedefaultseppunct}\relax
\EndOfBibitem
\bibitem[Aubry \emph{et~al.}(2008)Aubry, Singh, Janjua, and
  Nudurupati]{aubry2008micro}
N.~Aubry, P.~Singh, M.~Janjua and S.~Nudurupati, \emph{Proc. Natl. Acad. Sci.},
  2008, \textbf{105}, 3711--3714\relax
\mciteBstWouldAddEndPuncttrue
\mciteSetBstMidEndSepPunct{\mcitedefaultmidpunct}
{\mcitedefaultendpunct}{\mcitedefaultseppunct}\relax
\EndOfBibitem
\bibitem[Nishikawa \emph{et~al.}(2003)Nishikawa, Maenosono, Yamaguchi, and
  Okubo]{nishikawa2003self}
H.~Nishikawa, S.~Maenosono, Y.~Yamaguchi and T.~Okubo, \emph{J. Nanoparticle
  Res.}, 2003, \textbf{5}, 103--110\relax
\mciteBstWouldAddEndPuncttrue
\mciteSetBstMidEndSepPunct{\mcitedefaultmidpunct}
{\mcitedefaultendpunct}{\mcitedefaultseppunct}\relax
\EndOfBibitem
\bibitem[Gibbs \emph{et~al.}(1999)Gibbs, Kermasha, Alli, and
  Mulligan]{f1999encapsulation}
B.~F. Gibbs, S.~Kermasha, I.~Alli and C.~N. Mulligan, \emph{Int. J. Food Sci.
  Nutr.}, 1999, \textbf{50}, 213--224\relax
\mciteBstWouldAddEndPuncttrue
\mciteSetBstMidEndSepPunct{\mcitedefaultmidpunct}
{\mcitedefaultendpunct}{\mcitedefaultseppunct}\relax
\EndOfBibitem
\bibitem[Tsapis \emph{et~al.}(2002)Tsapis, Bennett, Jackson, Weitz, and
  Edwards]{tsapis2002trojan}
N.~Tsapis, D.~Bennett, B.~Jackson, D.~A. Weitz and D.~Edwards, \emph{Proc.
  Natl. Acad. Sci.}, 2002, \textbf{99}, 12001--12005\relax
\mciteBstWouldAddEndPuncttrue
\mciteSetBstMidEndSepPunct{\mcitedefaultmidpunct}
{\mcitedefaultendpunct}{\mcitedefaultseppunct}\relax
\EndOfBibitem
\bibitem[Bandi \emph{et~al.}(2011)Bandi, Tallinen, and
  Mahadevan]{bandi2011shock}
M.~Bandi, T.~Tallinen and L.~Mahadevan, \emph{EPL}, 2011, \textbf{96},
  36008\relax
\mciteBstWouldAddEndPuncttrue
\mciteSetBstMidEndSepPunct{\mcitedefaultmidpunct}
{\mcitedefaultendpunct}{\mcitedefaultseppunct}\relax
\EndOfBibitem
\bibitem[Knoche and Kierfeld(2015)]{knoche2015elasticity}
S.~Knoche and J.~Kierfeld, \emph{Langmuir}, 2015, \textbf{31}, 5364--5376\relax
\mciteBstWouldAddEndPuncttrue
\mciteSetBstMidEndSepPunct{\mcitedefaultmidpunct}
{\mcitedefaultendpunct}{\mcitedefaultseppunct}\relax
\EndOfBibitem
\bibitem[Cicuta and Vella(2009)]{cicuta2009granular}
P.~Cicuta and D.~Vella, \emph{Phys. Rev. Lett.}, 2009, \textbf{102},
  138302\relax
\mciteBstWouldAddEndPuncttrue
\mciteSetBstMidEndSepPunct{\mcitedefaultmidpunct}
{\mcitedefaultendpunct}{\mcitedefaultseppunct}\relax
\EndOfBibitem
\bibitem[Planchette \emph{et~al.}(2012)Planchette, Lorenceau, and
  Biance]{planchette2012surface}
C.~Planchette, E.~Lorenceau and A.-L. Biance, \emph{Soft Matter}, 2012,
  \textbf{8}, 2444--2451\relax
\mciteBstWouldAddEndPuncttrue
\mciteSetBstMidEndSepPunct{\mcitedefaultmidpunct}
{\mcitedefaultendpunct}{\mcitedefaultseppunct}\relax
\EndOfBibitem
\bibitem[Vella \emph{et~al.}(2006)Vella, Kim, Aussillous, and
  Mahadevan]{vella2006dynamics}
D.~Vella, H.-Y. Kim, P.~Aussillous and L.~Mahadevan, \emph{Phys. Rev. Lett.},
  2006, \textbf{96}, 178301\relax
\mciteBstWouldAddEndPuncttrue
\mciteSetBstMidEndSepPunct{\mcitedefaultmidpunct}
{\mcitedefaultendpunct}{\mcitedefaultseppunct}\relax
\EndOfBibitem
\bibitem[Rieser \emph{et~al.}(2015)Rieser, Arratia, Yodh, Gollub, and
  Durian]{rieser2015tunable}
J.~M. Rieser, P.~Arratia, A.~Yodh, J.~P. Gollub and D.~Durian, \emph{Langmuir},
  2015, \textbf{31}, 2421--2429\relax
\mciteBstWouldAddEndPuncttrue
\mciteSetBstMidEndSepPunct{\mcitedefaultmidpunct}
{\mcitedefaultendpunct}{\mcitedefaultseppunct}\relax
\EndOfBibitem
\bibitem[Harrington \emph{et~al.}(2019)Harrington, Liu, and
  Durian]{harrington2019machine}
M.~Harrington, A.~J. Liu and D.~J. Durian, \emph{Phys. Rev. E}, 2019,
  \textbf{99}, 022903\relax
\mciteBstWouldAddEndPuncttrue
\mciteSetBstMidEndSepPunct{\mcitedefaultmidpunct}
{\mcitedefaultendpunct}{\mcitedefaultseppunct}\relax
\EndOfBibitem
\bibitem[Rieser(2015)]{rieser2015deformation}
J.~M. Rieser, \emph{Deformation of two-dimensional amorphous granular
  packings}, University of Pennsylvania, 2015\relax
\mciteBstWouldAddEndPuncttrue
\mciteSetBstMidEndSepPunct{\mcitedefaultmidpunct}
{\mcitedefaultendpunct}{\mcitedefaultseppunct}\relax
\EndOfBibitem
\bibitem[Harrington \emph{et~al.}(2020)Harrington, Xiao, and
  Durian]{harrington2020stagnant}
M.~Harrington, H.~Xiao and D.~J. Durian, \emph{Granul. Matter}, 2020,
  \textbf{22}, 17\relax
\mciteBstWouldAddEndPuncttrue
\mciteSetBstMidEndSepPunct{\mcitedefaultmidpunct}
{\mcitedefaultendpunct}{\mcitedefaultseppunct}\relax
\EndOfBibitem
\bibitem[He \emph{et~al.}(2015)He, {\c{S}}enbil, and Dinsmore]{he2015measured}
W.~He, N.~{\c{S}}enbil and A.~Dinsmore, \emph{Soft Matter}, 2015, \textbf{11},
  5087--5094\relax
\mciteBstWouldAddEndPuncttrue
\mciteSetBstMidEndSepPunct{\mcitedefaultmidpunct}
{\mcitedefaultendpunct}{\mcitedefaultseppunct}\relax
\EndOfBibitem
\bibitem[Cook(1974)]{Cook1974}
R.~Cook, \emph{Concepts and Applications of Finite Element Analysis}, John
  Wiley and Sons, Inc., New York, 1974\relax
\mciteBstWouldAddEndPuncttrue
\mciteSetBstMidEndSepPunct{\mcitedefaultmidpunct}
{\mcitedefaultendpunct}{\mcitedefaultseppunct}\relax
\EndOfBibitem
\bibitem[Li \emph{et~al.}(2015)Li, Rieser, Liu, Durian, and
  Li]{li2015deformation}
W.~Li, J.~M. Rieser, A.~J. Liu, D.~J. Durian and J.~Li, \emph{Phys. Rev. E},
  2015, \textbf{91}, 062212\relax
\mciteBstWouldAddEndPuncttrue
\mciteSetBstMidEndSepPunct{\mcitedefaultmidpunct}
{\mcitedefaultendpunct}{\mcitedefaultseppunct}\relax
\EndOfBibitem
\bibitem[Le~Bouil \emph{et~al.}(2014)Le~Bouil, Amon, Sangleboeuf, Orain,
  B{\'e}suelle, Viggiani, Chasle, and Crassous]{le2014biaxial}
A.~Le~Bouil, A.~Amon, J.-C. Sangleboeuf, H.~Orain, P.~B{\'e}suelle,
  G.~Viggiani, P.~Chasle and J.~Crassous, \emph{Granular Matt.}, 2014,
  \textbf{16}, 1--8\relax
\mciteBstWouldAddEndPuncttrue
\mciteSetBstMidEndSepPunct{\mcitedefaultmidpunct}
{\mcitedefaultendpunct}{\mcitedefaultseppunct}\relax
\EndOfBibitem
\bibitem[Turnbull and Cohen(1961)]{Turnbull1961}
D.~Turnbull and M.~H. Cohen, \emph{J. Chem. Phys.}, 1961, \textbf{34},
  120--125\relax
\mciteBstWouldAddEndPuncttrue
\mciteSetBstMidEndSepPunct{\mcitedefaultmidpunct}
{\mcitedefaultendpunct}{\mcitedefaultseppunct}\relax
\EndOfBibitem
\bibitem[Schr\"{o}der-Turk \emph{et~al.}(2010)Schr\"{o}der-Turk, Mickel,
  Schröter, Delaney, Saadatfar, Senden, Mecke, and Aste]{SchroderTurkEPL2010}
G.~E. Schr\"{o}der-Turk, W.~Mickel, M.~Schröter, G.~W. Delaney, M.~Saadatfar,
  T.~J. Senden, K.~Mecke and T.~Aste, \emph{Europhys. Lett.}, 2010,
  \textbf{90}, 34001\relax
\mciteBstWouldAddEndPuncttrue
\mciteSetBstMidEndSepPunct{\mcitedefaultmidpunct}
{\mcitedefaultendpunct}{\mcitedefaultseppunct}\relax
\EndOfBibitem
\bibitem[Morse and Corwin(2014)]{MorseCorwinPRL2014}
P.~K. Morse and E.~I. Corwin, \emph{Phys. Rev. Lett.}, 2014, \textbf{112},
  115701\relax
\mciteBstWouldAddEndPuncttrue
\mciteSetBstMidEndSepPunct{\mcitedefaultmidpunct}
{\mcitedefaultendpunct}{\mcitedefaultseppunct}\relax
\EndOfBibitem
\bibitem[Cao \emph{et~al.}(2018)Cao, Li, Kou, Xia, Li, Chen, Xie, Xiao, Kob,
  Hong,\emph{et~al.}]{cao2018structural}
Y.~Cao, J.~Li, B.~Kou, C.~Xia, Z.~Li, R.~Chen, H.~Xie, T.~Xiao, W.~Kob, L.~Hong
  \emph{et~al.}, \emph{Nature Comm.}, 2018, \textbf{9}, 1--7\relax
\mciteBstWouldAddEndPuncttrue
\mciteSetBstMidEndSepPunct{\mcitedefaultmidpunct}
{\mcitedefaultendpunct}{\mcitedefaultseppunct}\relax
\EndOfBibitem
\bibitem[Cubuk \emph{et~al.}(2015)Cubuk, Schoenholz, Rieser, Malone, Rottler,
  Durian, Kaxiras, and Liu]{cubuk2015identifying}
E.~D. Cubuk, S.~S. Schoenholz, J.~M. Rieser, B.~D. Malone, J.~Rottler, D.~J.
  Durian, E.~Kaxiras and A.~J. Liu, \emph{Phys. Rev. Lett.}, 2015,
  \textbf{114}, 108001\relax
\mciteBstWouldAddEndPuncttrue
\mciteSetBstMidEndSepPunct{\mcitedefaultmidpunct}
{\mcitedefaultendpunct}{\mcitedefaultseppunct}\relax
\EndOfBibitem
\bibitem[Rieser \emph{et~al.}(2016)Rieser, Goodrich, Liu, and
  Durian]{RieserPRL2016}
J.~M. Rieser, C.~P. Goodrich, A.~J. Liu and D.~J. Durian, \emph{Phys. Rev.
  Lett.}, 2016, \textbf{116}, 088001\relax
\mciteBstWouldAddEndPuncttrue
\mciteSetBstMidEndSepPunct{\mcitedefaultmidpunct}
{\mcitedefaultendpunct}{\mcitedefaultseppunct}\relax
\EndOfBibitem
\bibitem[Lee \emph{et~al.}(2020)Lee, Buller, and
  Dalnoki-Veress]{lee2020cheerios}
C.~Lee, A.~Buller and K.~Dalnoki-Veress, \emph{Bull. Am. Phys. Soc.}, 2020,
  \textbf{65}, P15.00008\relax
\mciteBstWouldAddEndPuncttrue
\mciteSetBstMidEndSepPunct{\mcitedefaultmidpunct}
{\mcitedefaultendpunct}{\mcitedefaultseppunct}\relax
\EndOfBibitem
\bibitem[Boromand \emph{et~al.}(2019)Boromand, Signoriello, Lowensohn,
  Orellana, Weeks, Ye, Shattuck, and O'Hern]{boromand2019role}
A.~Boromand, A.~Signoriello, J.~Lowensohn, C.~S. Orellana, E.~R. Weeks, F.~Ye,
  M.~D. Shattuck and C.~S. O'Hern, \emph{Soft matter}, 2019, \textbf{15},
  5854--5865\relax
\mciteBstWouldAddEndPuncttrue
\mciteSetBstMidEndSepPunct{\mcitedefaultmidpunct}
{\mcitedefaultendpunct}{\mcitedefaultseppunct}\relax
\EndOfBibitem
\end{mcitethebibliography}
\bibliographystyle{rsc} 

\end{document}